\documentclass[a4paper]{raa_twocolumn}            

\usepackage{graphicx,times}             
\usepackage{natbib}
\usepackage{amssymb,amsmath}
\bibpunct{(}{)}{;}{a}{}{,}

\usepackage[a4paper=true,pagebackref=true,bookmarks=false]{hyperref}
\hypersetup{colorlinks = true, linkcolor = green, anchorcolor = red, citecolor = blue, filecolor = red, pagecolor = red, urlcolor = red}

\def\jref@jnl#1{{\rm#1\/}}
\def\actaa{\jref@jnl{Acta Astronomica}}
\def\aap{\jref@jnl{A\&A}}
\def\aapr{\jref@jnl{The Astronomy and Astrophysics Review}}
\def\aaps{\jref@jnl{Astronomy and Astrophysics Supplement Series}}
\def\aj{\jref@jnl{AJ}}
\def\apj{\jref@jnl{ApJ}}
\def\apjl{\jref@jnl{ApJL}}
\def\apjs{\jref@jnl{ApJS}}
\def\apss{\jref@jnl{Astrophysics and Space Science}}
\def\ao{\jref@jnl{Applied Optics}}
\def\araa{\jref@jnl{ARA\&A}}
\def\bain{\jref@jnl{BAN}}
\def\caa{\jref@jnl{Chinese Astronomy and Astrophysics}}
\def\cjaa{\jref@jnl{Chinese Journal of Astronomy and Astrophysics}}
\def\gca{\jref@jnl{Geochimica et Cosmochimica Acta}}
\def\jcp{\jref@jnl{Journal of Chemical Physics}}
\def\jqsrt{\jref@jnl{Journal of Quantitative Spectroscopy and Radiative Transfer}}
\def\mnras{\jref@jnl{MNRAS}}
\def\memras{\jref@jnl{Memoirs of the Royal Astronomical Society}}
\def\memsai{\jref@jnl{Memorie della Societa Astronomica Italiana}}
\def\na{\jref@jnl{New Astronomy}}
\def\nar{\jref@jnl{New Astronomy Reviews}}
\def\nat{\jref@jnl{Nature}}
\def\pasa{\jref@jnl{Publications of the Astronomical Society of Australia}}
\def\planss{\jref@jnl{Planetary and Space Science}}
\def\pasj{\jref@jnl{Publications of the Astronomical Society of Japan}}
\def\pasp{\jref@jnl{PASP}}
\def\physrep{\jref@jnl{Physics Reports}}
\def\pra{\jref@jnl{Physical Review A}}
\def\prd{\jref@jnl{Physical Review D}}
\def\pre{\jref@jnl{Physical Review E}}
\def\physrep{\jref@jnl{Physics Reports}}
\def\physscr{\jref@jnl{Physica Scripta}}
\def\qjras{\jref@jnl{Quarterly Journal of the Royal Astronomical Society}}
\def\rmxaa{\jref@jnl{Revista Mexicana de Astronomia y Astrofisica}}
\def\skytel{\jref@jnl{Sky and Telescope}}
\def\solphys{\jref@jnl{Solar Physics}}
\def\sovast{\jref@jnl{Soviet Astronomy}}
\def\ssr{\jref@jnl{Space Science Reviews}}
\def\zap{\jref@jnl{Zeitschrift fuer Astrophysik}}
\def\azh{\jref@jnl{Astronomicheskij Zhurnal}}
\def\procspie{\jref@jnl{Proc. SPIE}}

\newcommand{\MC}{\multicolumn}
\newcommand{\kms}{km\,s$^{-1}$}
\newcommand{\ms}{m\,s$^{-1}$}
\newcommand{\IC}{IC\,4662}
\DeclareRobustCommand{\ion}[2]{%
\relax\ifmmode
\ifx\testbx\f
{\mathrm{#1\,\textsc{#2}}}\else
{\mathrm{#1\,\mathsc{#2}}}\fi
\else\textup{#1\,{\mdseries\textsc{#2}}}%
\fi}

\begin{document}

    \title{Study of extragalactic \ion{H}{ii} regions with \'echelle spectroscopy: The A2 region in the irregular galaxy \IC}

   \volnopage{Vol.0 (20xx) No.0, 000--000}      
   \setcounter{page}{1}          

    \author{Alexei Y.\ Kniazev
        \inst{1,2,3}
    }

   \institute{%
            South African Astronomical Observatory, Cape Town, 7935, South Africa {\it a.kniazev@saao.nrf.ac.za}\\
        \and
            Southern African Large Telescope, Cape Town, 7935, South Africa\\
        \and
            Sternberg State Astronomical Institute, Moscow, Universitetsky ave., 13, Russia\\
\vs\no
   {\small Received~~2025 January 02; accepted~~20xx~~month day}}

\abstract{I present the results of \'echelle spectroscopy of a bright \ion{H}{ii} region in the irregular galaxy \IC\ and their comparison with results from long-slit spectroscopy of the same region. All observations were obtained with the standard spectrographs of the SALT telescope: (1) low and medium spectral resolution spectrograph RSS (R$\approx$800) and (2) \'echelle spectrograph HRS (R=16000-17000). In both types of data the intensities of most of the emission lines were measured and abundances of oxygen and N, Ne, S, Ar, Cl and Fe were determined as well as physical parameters of the \ion{H}{ii} region. The chemical abundances were obtained from both types of data with the T$_{\rm e}$-method. Abundances calculated from both types of data agree to within the cited uncertainties. {The analysis of the \'echelle data revealed three distinct kinematic subsystems within the studied \ion{H}{ii} region: a narrow component (NC, $\sigma \approx 12$~km/s), a broad component (BC, $\sigma \approx 40$~km/s), and a very broad component (VBC, $\sigma \approx 60-110$~km/s, detected only in the brightest emission lines). The elemental abundances for the NC and BC subsystems were determined using the T$_{\rm e}$-method. The velocity dispersion dependence on the ionisation potential of elements showed no correlation for the NC, indicating a well-mixed turbulent medium, while the BC exhibited pronounced stratification, characteristic of an expanding shell. Based on a detailed analysis of the kinematics and chemical composition, it was concluded that the BC is associated with the region surrounding a Wolf-Rayet star of spectral type WN7-8. The stellar wind from this WR star interacts with a shell ejected during an earlier evolutionary stage (either as a red supergiant or a luminous blue variable, LBV), which is enriched in nitrogen. These findings highlight the importance of high spectral resolution for detecting small-scale ($\sim25$~pc) chemical inhomogeneities and for understanding the feedback mechanisms of massive stars in low-metallicity environments.}
%
 \keywords{galaxies: abundances ---  galaxies: irregular --- galaxies: evolution  ---  galaxies: individual (\IC)}
}

   \authorrunning{A.\,Y.\,Kniazev}  
   \titlerunning{Spectroscopy \ion{H}{ii} region of the irregular galaxy \IC}  

   \maketitle


\begin{figure}
    \includegraphics[clip=,angle=0,width=0.48\textwidth]{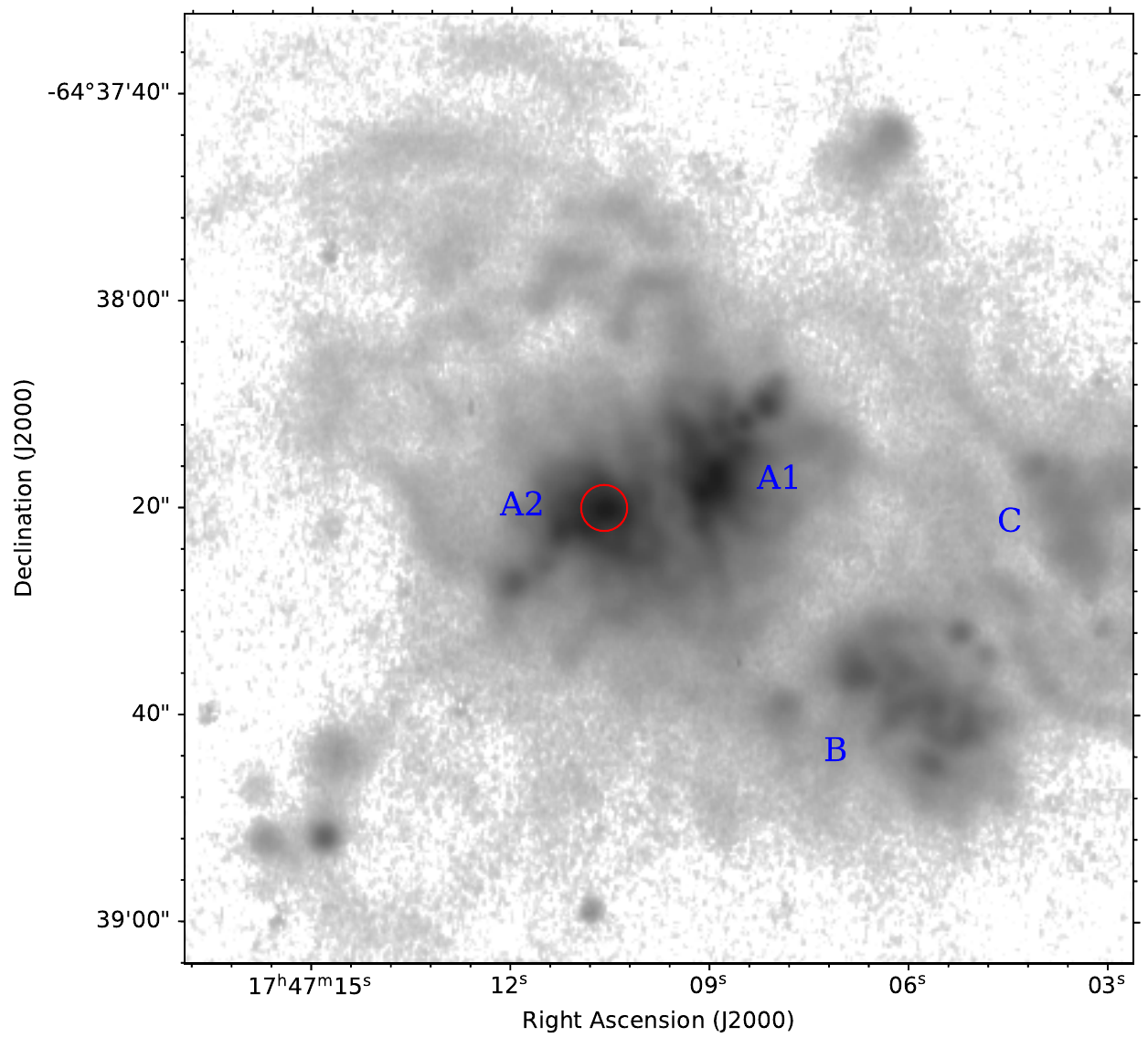}
    \includegraphics[clip=,angle=0,width=0.48\textwidth]{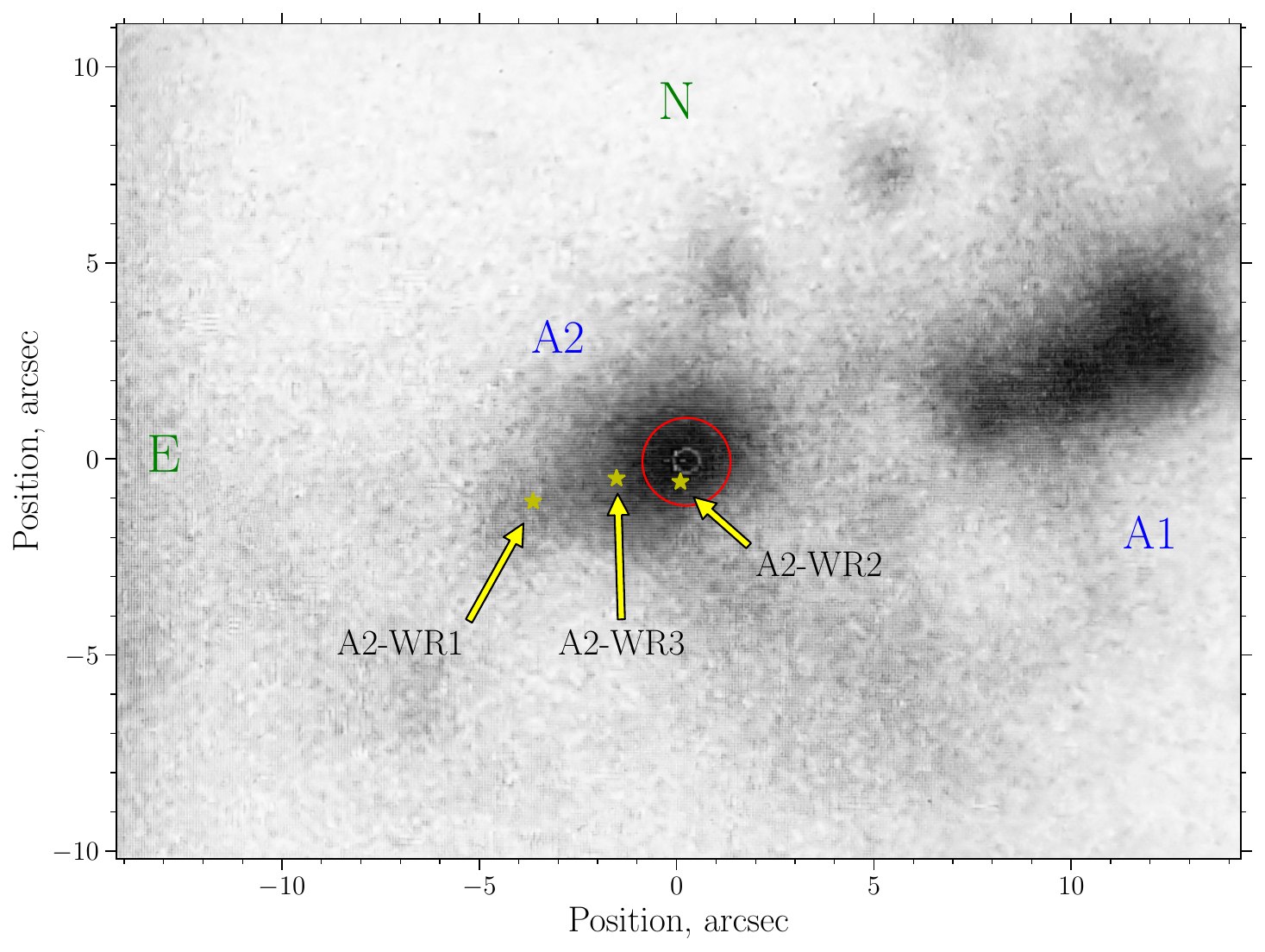}
    \caption{{\bf Left panel:} The central part of the H$\alpha$ \IC\ galaxy image.
        The image was taken with the ESO 2.2m telescope \citep{2007AstL...33..283K}.
        The image is displayed on a logarithmic scale to show the
        bright and faint H$\alpha$ regions at the same time.
        Darker objects indicate brighter sources of H$\alpha$ emission.
        The red circle shows the position of the HRS fibre ( 2.23\arcsec\ in diameter) 
        during the \'echelle observations.
        The spectral slit during observations with the RSS spectrograph 
        also passed through the same region.
        {\bf Right panel:} The image from the HRS guiding system 
        with size of 28.5$\times$21.3~arcsecs.
        The HRS guiding system has no filter and its maximum sensitivity is at $\sim6000$~\AA.
        The red circle is the exact position of the HRS fibre (diameter 2.23\arcsec)
        during the \'echelle observations.
        Yellow stars and arrows show positions and names of WR stars found 
        by \citet{2009A&A...499..455C}.
        At the assumed galaxy distance of 2.44 Mpc \citep{2007AstL...33..283K},
        1\arcsec\ = 11.8~pc.
        The naming system for \ion{H}{ii} regions is taken from \citet{1990A&A...234...99H}.
	\label{fig:IC4662_Ha}}
\end{figure}

\section{Introduction}

The study of extragalactic \ion{H}{ii} regions by spectroscopy usually involves
several fairly standard methods:
(1) long slit spectroscopy;
(2) multi-object spectroscopy;
(3) panoramic spectroscopy;
(4) Fabry-Perot spectroscopy;
(5) \'echelle spectroscopy.
Each of these methods has advantages and disadvantages, and each has a specific area of application. 
Only panoramic and \'echelle spectroscopy methods can be 
used for the high-resolution (R$>$10000) spectrophotometry of extragalactic \ion{H}{ii} regions
in a large spectral range.
In the case of panoramic spectroscopy it is usually necessary to use several
spectral configurations to cover a large spectral range, which requires 
stable weather conditions. 
This usually results in a spectral resolution of R$\le$5000
\citep{2006A&A...459...71I,2009MNRAS.398....2J,2013MNRAS.430.2097J,2013MNRAS.428...86J}.
Two problems, limited slit length for sky spectrum subtraction 
and blaze correction, appear when \'echelle spectroscopy is used. 
However, both these problems can be reasonably well resolved 
by using a fibre \'echelle spectrograph.

This paper presents the possibility of the study of chemical abundances in bright extragalactic 
\ion{H}{ii} regions using the high-resolution \'echelle spectrograph at the
Southern African Large Telescope \citep[SALT;][]{2006SPIE.6267E..0ZB,2006MNRAS.372..151O}.
Such a possibility will be demonstrated by comparison of chemical abundances 
in a bright \ion{H}{ii} region of the \IC\ galaxy, 
obtained by two spectroscopic methods -- long-slit and \'echelle.

\IC\ is a gas rich, metal poor, dwarf irregular galaxy at a 
distance of about 2.5 Mpc. It does not belong to any group and has no massive neighbors. 
However, the galaxy is actively star-forming and 
several giant \ion{H}{ii} regions are visible.
The radio and infrared data show that very strong star formation is going on
in the brightest \ion{H}{ii} regions
which is still weakly visible in the optical \citep{2003AJ....126..101J,2009Ap&SS.324..147G}.
The galaxy contains Wolf-Rayet (WR) stars \citep{2009A&A...499..455C}, and its
star formation history has been investigated by
\citet{2009ApJ...695..561M,2010ApJ...721..297M,2010ApJ...724...49M}.
\citet{2010MNRAS.407..113V} has examined in detail the \ion{H}{i} in \IC\ and has detected
a giant cloud of \ion{H}{i} around this galaxy.
The first detailed analysis of the chemical abundance 
of the \ion{H}{ii} regions of this galaxy was done
by \citet{1990A&A...234...99H}, where it was shown that the two brightest \ion{H}{ii} regions have very 
similar chemical abundances.
In \citet{2001A&A...369..797H} it was shown that there is a fainter \ion{H}{ii} region
on the edge of \IC, which has a metallicity significantly different {(0.5~dex)} from the metallicities
of the central \ion{H}{ii} region.

This article is organised as follows:
Section~\ref{txt:Obs_and_Red} describes observations and data reduction.
Section~\ref{txt:analysis} briefly describes the methodology of the analysis.
Section~\ref{txt:results} presents results of our analysis,
and their discussion is presented in Section~\ref{txt:disc}.

\begin{figure*}
    \centering{
    \includegraphics[clip=,angle=0,width=15.0cm]{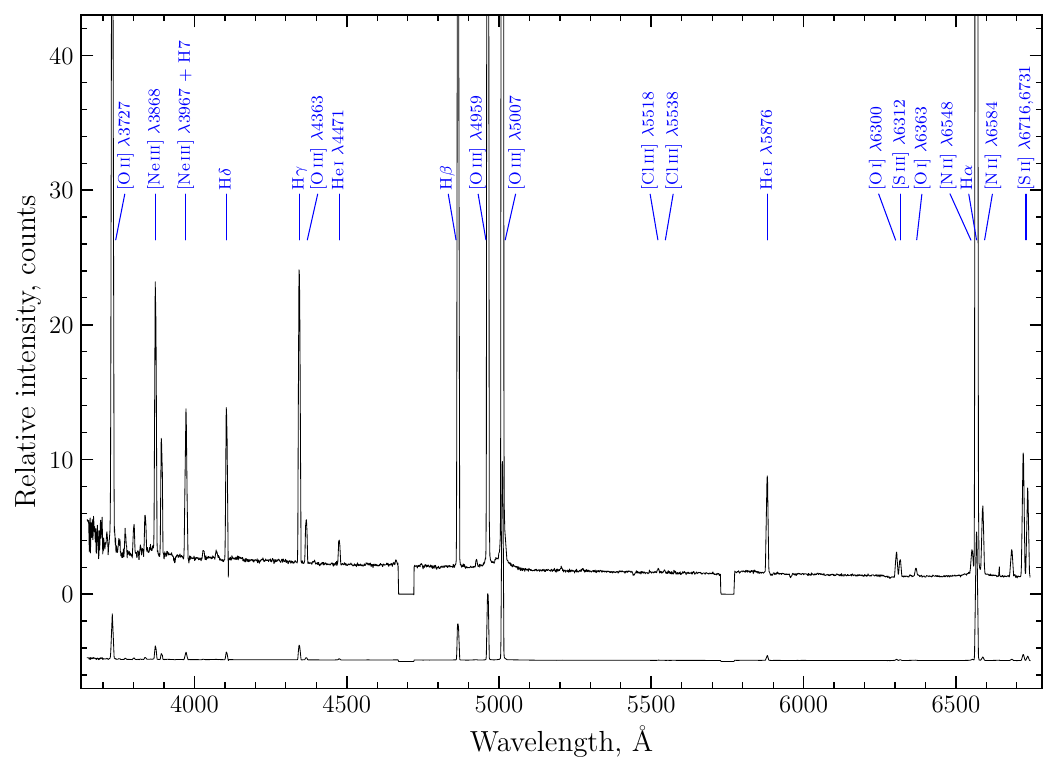}
    \caption{One dimensional spectrum of the A2 region obtained 
        with a long slit in the spectral region 3650--6750~\AA.
        All major emission lines are shown and their
        measurements are given in Table~\ref{tab:Lines}.
        The spectrum at the bottom is scaled and
        shifted to show the relative intensities of strong lines.
        Two gaps between the CCDs in the mosaic are obvious, with no data there.
        \label{fig:IC4662_A2_long}}
}
\end{figure*}

\section{Spectral observations and data reduction}
\label{txt:Obs_and_Red}

\subsection{Long-slit observations}
\label{txt:long-slit}

The long-slit spectral observations of the \ion{H}{ii} region A2 in the \IC\ galaxy
(see Figure~\ref{fig:IC4662_Ha})
were carried out during SALT commissioning on April 25th, 2006
with the Robert Stobie Spectrograph, 
\citep[RSS hereafter;][]{2003SPIE.4841.1463B, 2003SPIE.4841.1634K} 
which is a spectrograph of low and medium spectral resolution.
Observations were done with an image quality of 1\farcs4.
A mosaic of three CCDs is used as the detector on the RSS,
with the total mosaic size being 6144$\times$4096 pixels.
The pixel size along the slit is 0\farcs129 and the total slit length is 8\arcmin.
A binning factor of 2 was used during observations of \IC\ 
resulting in a pixel size along the slit of 0\farcs258.
The observations were carried out with a spectral configuration covering the spectral range 
3650--6750~\AA.
The Volume Phase Holographic (VPH) grating, PG900, was used
with a final reciprocal dispersion of about 0.97\AA\ per pixel. 
A slit width of 1\farcs5 resulted in a spectral resolution of { $R\sim800$ (FWHM$\sim4.8$~\AA)}.
Two 600~sec exposures were made, followed by a reference spectrum of a CuAr lamp
and spectral flat-fields. 
Zero level (bias) images were also taken
for a standard processing of the two-dimensional spectra.
In order to get the correct relative energy distribution,
spectra of the spectrophotometric standard EG\,21 \citep{1992PASP..104..533H,1994PASP..106..566H} 
were obtained during astronomical twilight.
SALT is a telescope with a variable pupil, so that the illuminating beam changes continuously during 
the observations. This makes absolute flux calibration impossible even when using 
spectrophotometric standards.
However, the relative energy distributions are very accurate, 
especially as the SALT has an atmospheric dispersion compensator (ADC).

The primary data reduction was done using the standard SALT pipeline \citep{2010SPIE.7737E..25C},
and the following preliminary and spectral data reduction 
were done using an RSS spectral pipeline \cite{2021AstBu..K}.
For the analysis presented in this paper,
a one-dimensional spectrum of the A2 region (9 pixels or 2\farcs3) was extracted from the 
two-dimensional reduced spectrum.
This spectrum is shown in Figure~\ref{fig:IC4662_A2_long},
where the most important emission lines are labelled.

\begin{figure*}
    \centering{
    \includegraphics[clip=,angle=0,width=0.9\textwidth]{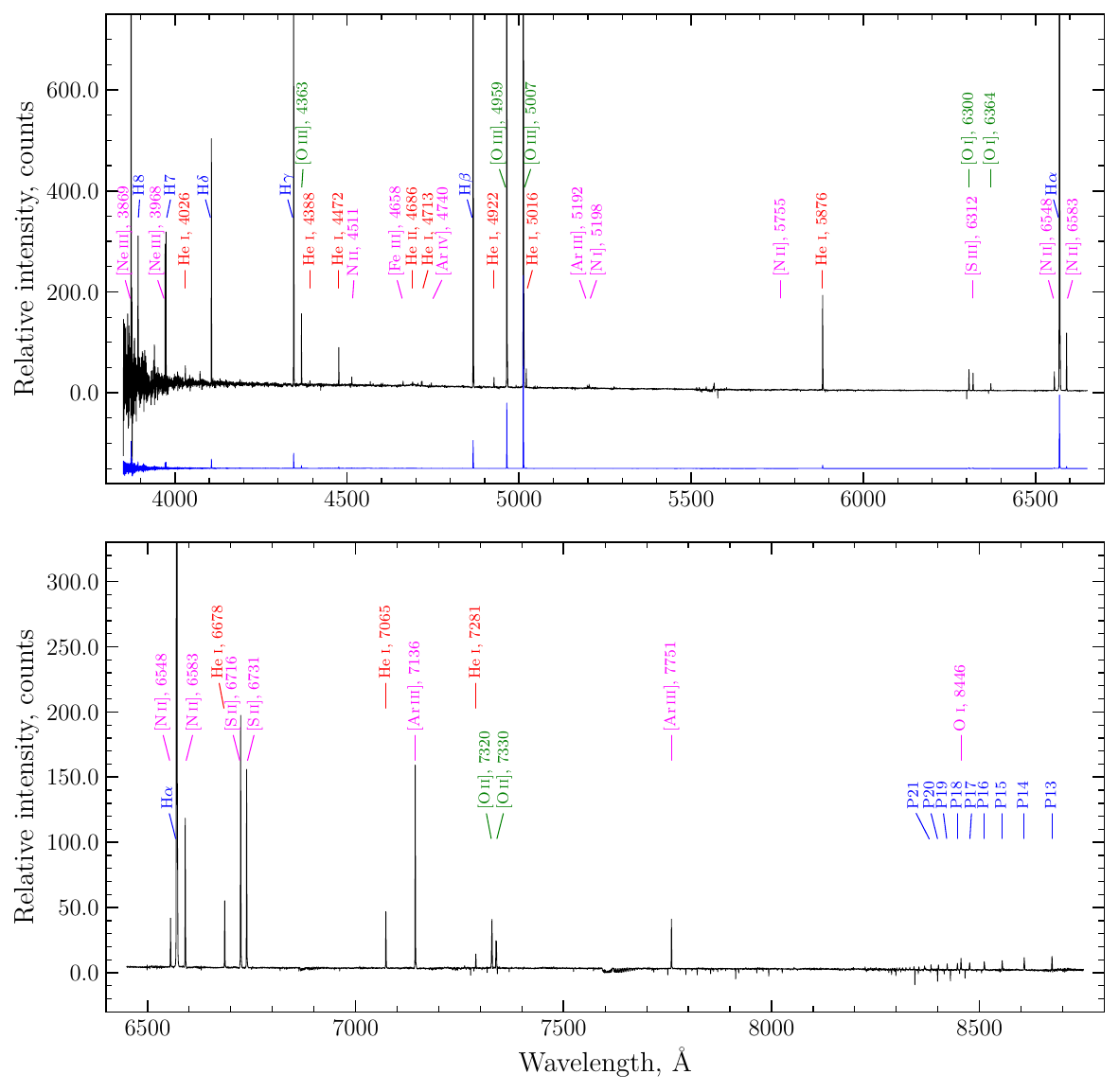}
    \caption{The final \'echelle spectrum of the A2 region shown in two panels
        (74 spectral orders were merged, and corrected for the sensitivity curve).
        All the main emission lines are shown and their intensities are presented
        in Table~\ref{tab:Lines}.
        In the upper panel, the spectrum is again displayed, compressed vertically by a factor of 20 and shifted downwards
         to show the relative strengths of the strongest lines.
        \label{fig:IC4662_A2_ech}}
}
\end{figure*}

\subsection{\'Echelle observations}
\label{txt:echelle}

Spectral observations of the A2 region of the galaxy \IC\ employing the
High Resolution Fibre \'echelle  Spectrograph
\citep[HRS;][]{2008SPIE.7014E..0KB,2010SPIE.7735E..4FB,2012SPIE.8446E..0AB,2014SPIE.9147E..6TC}
were carried out on 11 October 2017.
A single exposure of 2400~sec duration was performed with seeing of 1\farcs5.
The HRS is a thermostabilised double-beam \'echelle spectrograph, with the entire optical part
housed in a vacuum to reduce temperature and mechanical influences.
The blue arm of the spectrograph covers the spectral range of 3735--5580~\AA,
and the red arm covers the spectral range of 5415--8870~\AA, respectively.
The spectrograph can be used in low (LR, R$\approx$14,000--15,000)
medium (MR, R$\approx$40,000--43,000) and high (HR, R$\approx$67,000--74,000) resolution modes
and is equipped with two fibres (object and sky fibres) for each mode.
During observations of the A2 region the LR mode was used
with fibres of 2\farcs23 diameter.
Both CCD detectors, for the blue and red arms,
were read out with a binning of 1$\times$1.

All standard calibrations for HRS are done once a week, which is technically
sufficient to achieve a spectral accuracy of 400-500~\ms\ for the LR mode.
The current HRS calibration plan includes: 
(1) three flat-field spectra to find the positions of all spectral \'echelle 
orders and correct for the effect of the brightness distribution along each \'echelle order 
(blaze correction) and 
(2) one (Th+Ar) lamp spectrum for wavelength calibration.
Also, 11 bias images are taken at the beginning of each observational night for
subsequent zero-offset accounting.
Once a week, one velocity standard is observed in all modes to check the accuracy of the wavelength calibration. 
All velocity standards have a velocity accuracy better than 20~\ms\ 
\citep[see for details][]{2019AstBu..74..208K}.
An HRS spectrophotometric standard is observed once per month.
All HRS spectrophotometric standards have spectral distributions known in steps of 3--4~\AA.

The primary HRS data reduction was performed automatically using the
SALT standard pipeline \citep{2010SPIE.7737E..25C} 
and the following \'echelle data reduction was done using the HRS pipeline 
described in detail in \citet{2016MNRAS.459.3068K,2019AstBu..74..208K}.
It should be noted here that the algorithm for constructing a two-dimensional dispersion 
curve is the same for all HRS modes, but the optical path of the light in each mode is slightly
different and the wavelength calibration result is different for different modes.
Initial and final wavelengths for each \'echelle order in the LR mode,
the number of identified lines used and the final solution accuracy for each order
are given in Tables~\ref{tab:Orders_blue} and \ref{tab:Orders_red}.
An improved version of the pipeline has been used for HRS data since September 2020:
(1) the removal of scattered light between orders has been significantly improved,
and (2) stability of the wavelength extraction and calibration
for the five bluer \'echelle orders up to wavelength $\approx3732$~\AA\ has been improved.
Thus, 41 \'echelle orders are now extracted and calibrated in the blue region.

\begin{figure}
    \includegraphics[clip=,angle=-90,width=8.0cm]{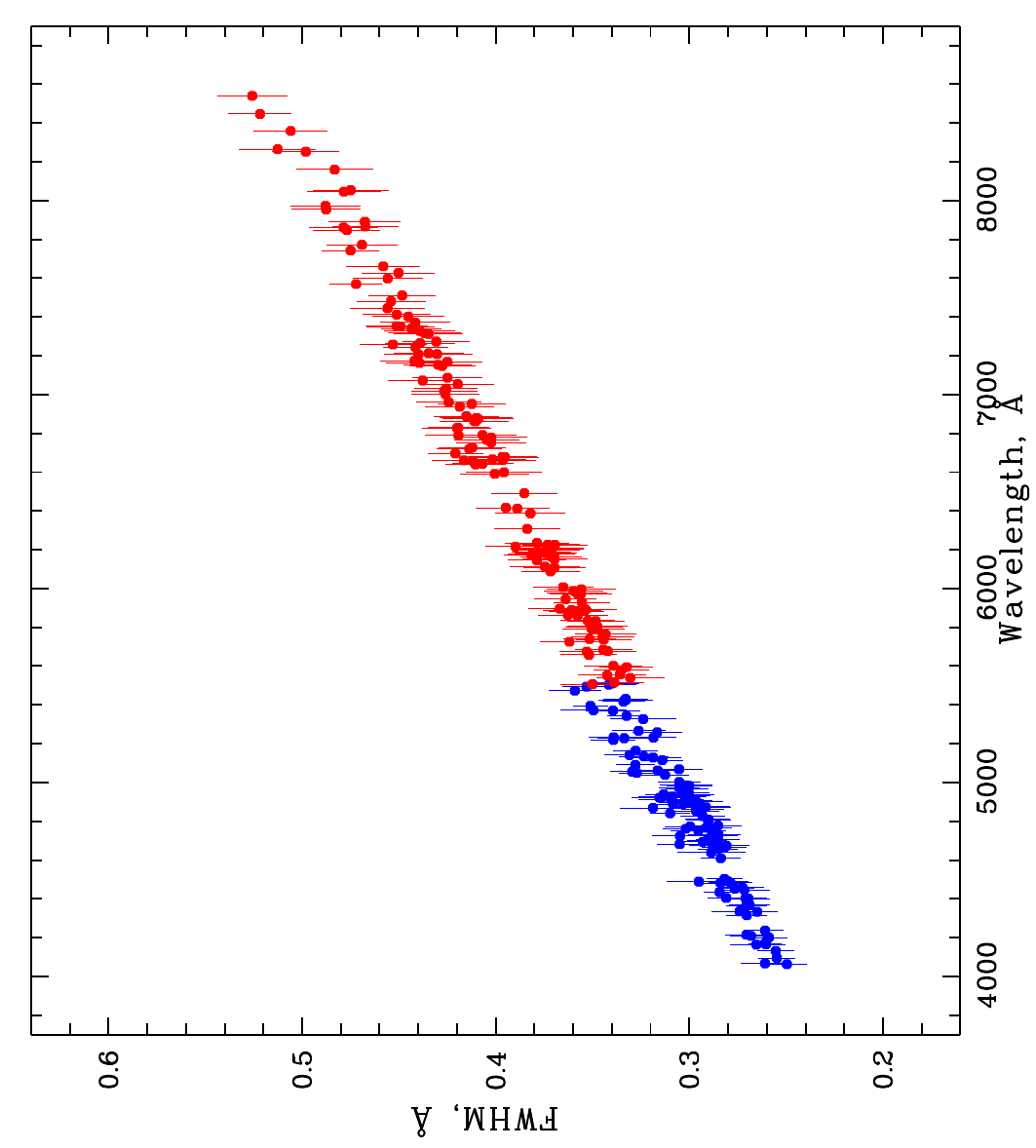}
    \caption{%
        { Wavelength dependence of FWHM} for the blue and red arms for the LR mode.
        \label{fig:FWHM}}
\end{figure}

{ FWHM measured} over all lines found in the reduced 
reference spectra varies from $\approx$0.25~\AA\ to $\approx$0.34~\AA\
for the blue arm spectrum and from
$\approx$0.33~\AA\ to $\approx$0.53~\AA\ for the red arm spectrum.
The change in the half-width of the instrumental profile as a function of wavelength
is shown in Figure~\ref{fig:FWHM}.
The behaviour of the instrumental profile is well approximated by a first order polynomial
and can be written in the form:
\begin{equation}
    \label{eq:blue}
    {\rm FWHM}(\lambda) =6.4148\cdot10^{-5} \cdot \lambda - 9.1383\cdot10^{-3},
\end{equation}
for the blue spectrum in the spectral region 3715--5580~\AA\ giving an accuracy of 0.007~\AA, and
\begin{equation}
    \label{eq:red}
    {\rm FWHM}(\lambda) = 5.9818\cdot10^{-5} \cdot \lambda + 4.9624\cdot10^{-3},
\end{equation}
for the red spectrum in the spectral region 5415--8870~\AA\ giving an accuracy of 0.006~\AA.

\begin{figure}
    \includegraphics[clip=,angle=-90,width=8.0cm]{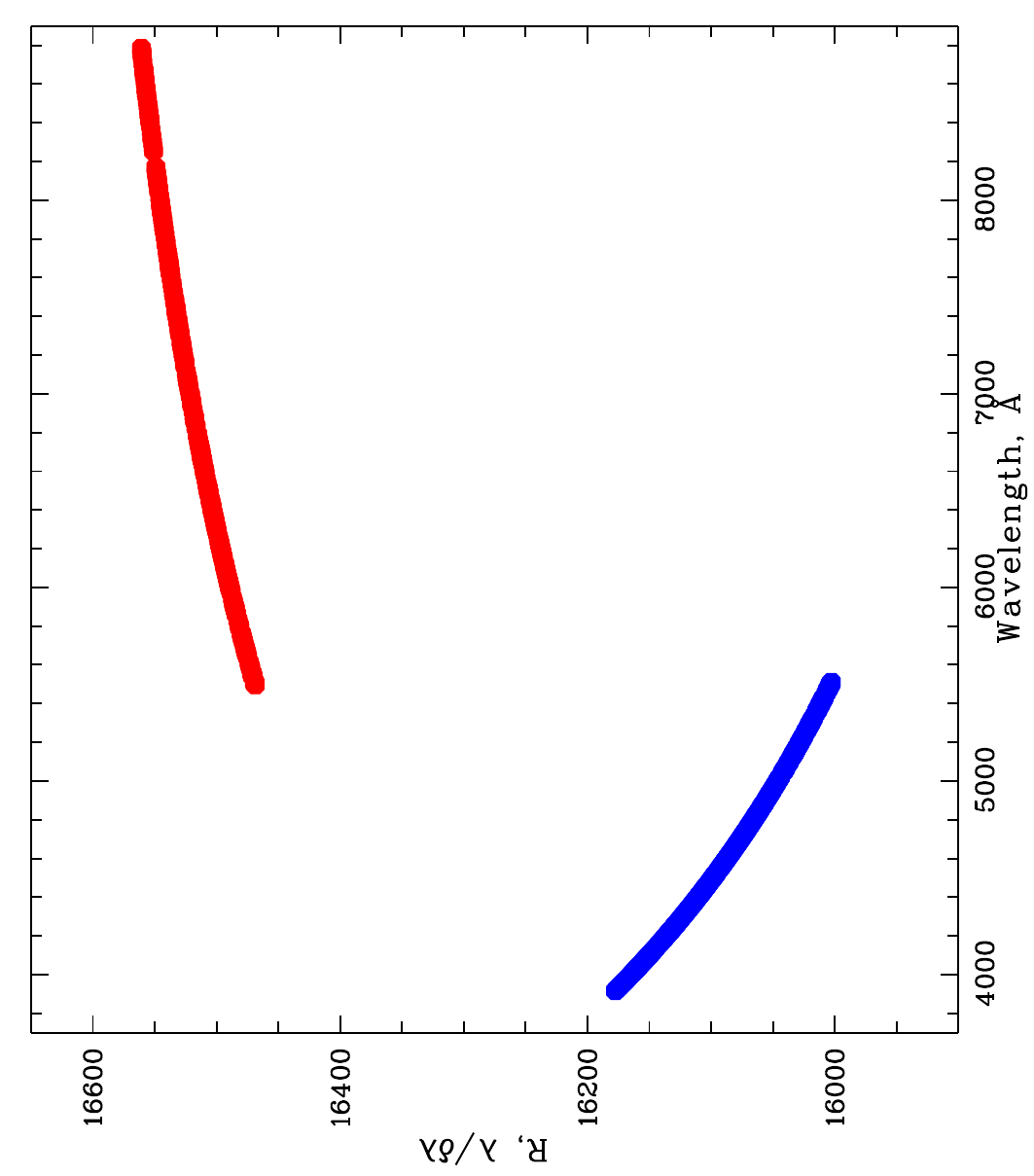}
    \caption{%
        Spectral resolution, $\rm R = \lambda/\delta\lambda$, as a function of wavelength
        for the blue and red arms, for the LR mode.
        \label{fig:R}}
\end{figure}

The spectral resolution $\rm R = \lambda/\delta\lambda$ as a function of wavelength
is shown in Figure~\ref{fig:R}.
As can be seen from the figure, the resulting resolution range, R=16,000--16,600, 
is slightly  higher than published for the spectrograph, viz., 
R=14,000--15,000\footnote{\url{http://pysalt.salt.ac.za/proposal_calls/current/ProposalCall.html}},
but the difference is no more than 10 \%.
The final \'echelle spectrum of the A2 region, comprising 74 spectral orders merged and corrected
for the sensitivity curve, is shown in Figure~\ref{fig:IC4662_A2_ech}.
The most important emission lines visible on this spectrum are indicated.

\begin{table*}
    \centering{
    \caption{Spectral orders for the HRS blue arm in LR mode\label{tab:Orders_blue}}
    \begin{tabular}{r|c|c|c|c|c|c|c|c|c}
        \hline
        \#       &Spectral&Spectral& \MC{2}{c}{Object Fiber}   & RMS      &Spectral& \MC{2}{c}{Sky Fiber}  & RMS  \\
                 &Order   &Lines   &$\lambda_0$&$\lambda_{end}$&          &Lines   &$\lambda_0$&$\lambda_{end}$&  \\
                 &        &        &  (\AA)    &    (\AA)      & (\AA)    &        &  (\AA)    &   (\AA)       & (\AA)\\
        \hline
        1   &  124  &      1  & 3732.48   & 3782.27 & 0.05051  &     1 & 3732.49  &  3782.26 &  0.04968 \\
        2   &  123  &      1  & 3762.92   & 3812.99 & 0.15497  &     1 & 3762.92  &  3812.98 &  0.15412 \\
        3   &  122  &      5  & 3792.01   & 3844.50 & 0.01361  &     5 & 3792.01  &  3844.50 &  0.01361 \\
        4   &  121  &      3  & 3825.28   & 3875.94 & 0.08576  &     3 & 3825.28  &  3875.94 &  0.08487 \\
        5   &  120  &      5  & 3857.32   & 3908.02 & 0.00446  &     4 & 3857.20  &  3908.31 &  0.01274 \\
        6   &  119  &      8  & 3889.94   & 3940.96 & 0.00755  &     7 & 3889.88  &  3941.04 &  0.00703 \\
        7   &  118  &      8  & 3922.74   & 3974.40 & 0.00443  &    12 & 3923.06  &  3974.35 &  0.01131 \\
        8   &  117  &      9  & 3956.49   & 4008.26 & 0.00674  &     8 & 3956.91  &  4008.34 &  0.00581 \\
        9   &  116  &      9  & 3990.43   & 4042.80 & 0.00858  &    10 & 3990.75  &  4042.91 &  0.00777 \\
        10  &  115  &     15  & 4025.33   & 4077.92 & 0.00481  &    15 & 4025.42  &  4078.04 &  0.00824 \\
        11  &  114  &     16  & 4060.75   & 4113.65 & 0.00337  &    17 & 4060.78  &  4113.77 &  0.00749 \\
        12  &  113  &     13  & 4096.71   & 4149.99 & 0.00607  &    12 & 4096.82  &  4150.10 &  0.00392 \\
        13  &  112  &      8  & 4133.33   & 4186.99 & 0.00519  &    10 & 4133.42  &  4187.10 &  0.00705 \\
        14  &  111  &     15  & 4170.67   & 4224.69 & 0.00474  &    16 & 4170.79  &  4224.79 &  0.00664 \\
        15  &  110  &     15  & 4208.60   & 4263.07 & 0.00530  &    16 & 4208.72  &  4263.19 &  0.00449 \\
        16  &  109  &     19  & 4247.25   & 4302.13 & 0.00722  &    17 & 4247.37  &  4302.24 &  0.00750 \\
        17  &  108  &     11  & 4286.49   & 4342.04 & 0.01378  &    11 & 4286.57  &  4342.18 &  0.00920 \\
        18  &  107  &     13  & 4326.76   & 4382.45 & 0.00591  &    16 & 4326.87  &  4382.57 &  0.00704 \\
        19  &  106  &     16  & 4367.61   & 4423.76 & 0.00873  &    14 & 4367.73  &  4423.89 &  0.00397 \\
        20  &  105  &     16  & 4409.27   & 4465.84 & 0.00677  &    14 & 4409.41  &  4465.94 &  0.00694 \\
        21  &  104  &     13  & 4451.66   & 4508.74 & 0.00692  &    14 & 4451.79  &  4508.86 &  0.00752 \\
        22  &  103  &     19  & 4494.93   & 4552.51 & 0.00939  &    17 & 4495.07  &  4552.60 &  0.00584 \\
        23  &  102  &     17  & 4539.06   & 4597.08 & 0.00800  &    15 & 4539.16  &  4597.22 &  0.00960 \\
        24  &  101  &     17  & 4584.05   & 4642.54 & 0.00881  &    15 & 4584.15  &  4642.74 &  0.00763 \\
        25  &  100  &     19  & 4629.93   & 4688.96 & 0.00775  &    18 & 4630.06  &  4689.08 &  0.00783 \\
        26  &   99  &     18  & 4676.75   & 4736.16 & 0.00843  &    16 & 4676.84  &  4736.44 &  0.00682 \\
        27  &   98  &     14  & 4724.46   & 4784.57 & 0.00842  &    12 & 4724.58  &  4784.70 &  0.00724 \\
        28  &   97  &     17  & 4773.15   & 4833.88 & 0.00987  &    17 & 4773.33  &  4834.00 &  0.00858 \\
        29  &   96  &     19  & 4822.94   & 4884.21 & 0.00677  &    19 & 4823.06  &  4884.34 &  0.00615 \\
        30  &   95  &     19  & 4873.73   & 4935.60 & 0.00854  &    20 & 4873.86  &  4935.73 &  0.00888 \\
        31  &   94  &     15  & 4925.58   & 4988.09 & 0.00907  &    15 & 4925.75  &  4988.21 &  0.00750 \\
        32  &   93  &      8  & 4978.59   & 5041.46 & 0.00803  &     8 & 4978.72  &  5041.47 &  0.00999 \\
        33  &   92  &     13  & 5032.68   & 5096.55 & 0.00598  &     7 & 5031.72  &  5096.63 &  0.00518 \\
        34  &   91  &     15  & 5088.03   & 5152.46 & 0.00802  &    14 & 5088.16  &  5152.61 &  0.00687 \\
        35  &   90  &     17  & 5144.58   & 5209.67 & 0.00677  &    15 & 5144.68  &  5209.84 &  0.00741 \\
        36  &   89  &     16  & 5202.33   & 5268.20 & 0.00882  &    15 & 5202.45  &  5268.36 &  0.00836 \\
        37  &   88  &     17  & 5261.44   & 5328.06 & 0.00714  &    15 & 5261.53  &  5328.22 &  0.00629 \\
        38  &   87  &     15  & 5321.89   & 5389.38 & 0.01154  &    12 & 5322.05  &  5389.52 &  0.00982 \\
        39  &   86  &     16  & 5383.82   & 5451.94 & 0.01082  &    15 & 5383.97  &  5452.09 &  0.00978 \\
        40  &   85  &     16  & 5447.09   & 5516.12 & 0.00785  &    15 & 5447.24  &  5516.27 &  0.00640 \\
        41  &   84  &      6  & 5511.89   & 5581.96 & 0.00486  &     7 & 5512.05  &  5582.06 &  0.00190 \\
        \hline
    \end{tabular}
}
\end{table*}
%
\begin{table*}
    \centering{
    \caption{Spectral orders for the HRS red arm in LR mode\label{tab:Orders_red}}
    \begin{tabular}{r|c|c|c|c|c|c|c|c|c}
        \hline
        \#       &Spectral&Spectral& \MC{2}{c}{Object Fiber}   & RMS      &Spectral& \MC{2}{c}{Sky Fiber}  & RMS  \\
        &Order   &Lines   &$\lambda_0$&$\lambda_{end}$&          &Lines   &$\lambda_0$&$\lambda_{end}$&  \\
        &        &        &  (\AA)    &    (\AA)      & (\AA)    &        &  (\AA)    &   (\AA)       & (\AA)\\
        \hline
        1   &   85   &    4   & 5413.92   & 5531.27 & 0.00378  &   5   & 5413.71   &  5531.16 &   0.00383     \\
        2   &   84   &   11   & 5478.41   & 5597.11 & 0.00510  &   9   & 5478.21   &  5596.98 &   0.00621     \\
        3   &   83   &    7   & 5544.45   & 5664.53 & 0.00238  &  10   & 5544.26   &  5664.39 &   0.00400     \\
        4   &   82   &   13   & 5612.10   & 5733.59 & 0.00388  &  14   & 5611.91   &  5733.44 &   0.00522     \\
        5   &   81   &   12   & 5681.42   & 5804.36 & 0.00450  &  12   & 5681.24   &  5804.19 &   0.00590     \\
        6   &   80   &   10   & 5752.48   & 5876.89 & 0.00246  &   9   & 5752.30   &  5876.72 &   0.00358     \\
        7   &   79   &   12   & 5825.33   & 5951.26 & 0.00265  &  15   & 5825.16   &  5951.08 &   0.00395     \\
        8   &   78   &   15   & 5900.05   & 6027.54 & 0.00465  &  16   & 5899.88   &  6027.34 &   0.00430     \\
        9   &   77   &   21   & 5976.71   & 6105.79 & 0.00449  &  22   & 5976.54   &  6105.59 &   0.00518     \\
        10  &   76   &   18   & 6055.39   & 6186.11 & 0.00728  &  21   & 6055.22   &  6185.90 &   0.00575     \\
        11  &   75   &   19   & 6136.17   & 6268.56 & 0.00509  &  24   & 6136.00   &  6268.35 &   0.00539     \\
        12  &   74   &   17   & 6219.13   & 6353.24 & 0.00525  &  21   & 6218.96   &  6353.03 &   0.00659     \\
        13  &   73   &   14   & 6304.36   & 6440.24 & 0.00341  &  16   & 6304.19   &  6440.02 &   0.00527     \\
        14  &   72   &    9   & 6391.96   & 6529.66 & 0.00414  &  17   & 6391.78   &  6529.44 &   0.00382     \\
        15  &   71   &   13   & 6482.02   & 6621.60 & 0.00844  &  13   & 6481.85   &  6621.37 &   0.00803     \\
        16  &   70   &   17   & 6574.66   & 6716.16 & 0.00534  &  23   & 6574.48   &  6715.93 &   0.00613     \\
        17  &   69   &   13   & 6669.98   & 6813.46 & 0.00640  &  15   & 6669.80   &  6813.23 &   0.00559     \\
        18  &   68   &   11   & 6768.10   & 6913.63 & 0.00463  &  15   & 6767.92   &  6913.40 &   0.00427     \\
        19  &   67   &   14   & 6869.15   & 7016.78 & 0.00600  &  16   & 6868.96   &  7016.55 &   0.00625     \\
        20  &   66   &   13   & 6973.26   & 7123.06 & 0.00483  &  17   & 6973.07   &  7122.84 &   0.00502     \\
        21  &   65   &   16   & 7080.57   & 7232.62 & 0.00498  &  18   & 7080.38   &  7232.39 &   0.00553     \\
        22  &   64   &   18   & 7191.24   & 7345.59 & 0.00437  &  18   & 7191.04   &  7345.36 &   0.00626     \\
        23  &   63   &   17   & 7305.42   & 7462.16 & 0.00552  &  17   & 7305.21   &  7461.93 &   0.00514     \\
        24  &   62   &   11   & 7423.27   & 7582.48 & 0.00518  &  12   & 7423.06   &  7582.25 &   0.00672     \\
        25  &   61   &   14   & 7544.99   & 7706.75 & 0.00548  &  13   & 7544.77   &  7706.52 &   0.00495     \\
        26  &   60   &    7   & 7670.77   & 7835.16 & 0.00420  &   7   & 7670.54   &  7834.94 &   0.00458     \\
        27  &   59   &   13   & 7800.80   & 7967.93 & 0.00695  &  13   & 7800.57   &  7967.70 &   0.00583     \\
        28  &   58   &   13   & 7935.32   & 8105.28 & 0.00803  &  12   & 7935.09   &  8105.05 &   0.00763     \\
        29  &   57   &    7   & 8074.56   & 8247.45 & 0.00764  &   5   & 8074.32   &  8247.22 &   0.00753     \\
        30  &   56   &    6   & 8218.76   & 8394.70 & 0.00537  &   6   & 8218.52   &  8394.46 &   0.00569     \\
        31  &   55   &    7   & 8368.21   & 8547.31 & 0.00679  &   8   & 8367.96   &  8547.06 &   0.00356     \\
        32  &   54   &    8   & 8523.18   & 8705.57 & 0.00598  &   7   & 8522.93   &  8705.32 &   0.00681     \\
        33  &   53   &    7   & 8684.01   & 8869.81 & 0.00418  &   3   & 8683.75   &  8869.54 &   0.00216     \\
        \hline
    \end{tabular}
}
\end{table*}

\section{Physical Conditions and Determination of Heavy-Element Abundances}
\label{txt:analysis}

All emission lines for both types of data have been measured using programs described in detail 
in \cite{2004ApJS..153..429K}.
The programs can measure the emission line intensities in two ways: 
(1) as the total flux exceeding a fitted continuum at some wavelength interval and then the shape 
of the lines is not taken into account,
or (2) by fitting a Gaussian function (or number of Gaussians) to the observed flux distribution 
exceeding the fitted continuum in some wavelength interval.
Usually, method (1) is used except for situations when two measured lines are close to each other
and their fluxes overlap and it is necessary to separate them.
It is important to estimate the total measurement errors of the line fluxes as a combination of the 
errors:
(1) errors of continuum fitting,
(2) errors due to the Poisson statistics of photons in lines, and 
(3) errors of the spectral sensitivity curve.
For the described observations the error (3) was about 2\% for both types of data.
All error components were summed quadratically and the total error
in the line intensities was used by the programs to calculate the error of the output physical parameters 
and chemical composition.
It should be noted here that the original set of programs described in \cite{2004ApJS..153..429K}
was created in the standard astronomical data processing system, \textsc{MIDAS}, which, since April 2012, is no longer
officially supported by the European Southern Observatory (ESO).
For this reason the author has rewritten all the relevant \textsc{MIDAS} programs in Python.

The analysis of the emission spectra was carried out within the classical two-zone model of the \ion{H}{ii} region. This methodology is described in detail in \citet{2004ApJS..153..429K,2005AJ....130.1558K,2008MNRAS.388.1667K,2012AstL...38..707K}. The chemical element contents of O, N, Ne, S, Ar, Cl and Fe as well as physical parameters such as the electron temperature ($\rm T_e$) and electron density ($\rm N_e$) were determined using this methodology. All the above parameters and their errors were determined taking into account the errors of the emission line intensities measured in the previous step. The spectral data for the A2 region in \IC\ covered the whole required spectral range in both cases, although in the case of the long-slit spectra the lines of \ion{He}{ii}~$\lambda$4686~\AA\ and [\ion{N}{ii}]~$\lambda$5755~\AA\ were lost due to the gaps between the CCDs in the RSS mosaic. For this reason two independent electron temperatures were used in the calculations based on the \'echelle data: $\rm T_e$(\ion{O}{iii}) for the hot zone and $\rm T_e$(\ion{N}{ii}) for the cold zone. The former was calculated from the line ratios of [\ion{O}{iii}]~$\lambda\lambda$4363,4959,5007~\AA, and the second from the line ratios of [\ion{N}{ii}]~$\lambda\lambda$5755,6548,6584~\AA. The total oxygen abundance was calculated as:
\begin{equation}
    \frac{O}{H} = \frac{O^{+}}{H} + \frac{O^{++}}{H} + \frac{O^{+++}}{H}
\end{equation}
In the case of the long-slit data, the cold zone temperature was calculated from an approximation based on the hot zone temperature using equations from \citet{2006A&A...448..955I}, and the total oxygen abundance was calculated without $\rm O^{+++}/H$, as this requires the presence of the \ion{He}{ii}~$\lambda$4686~\AA\ line. However, it should be noted here that the measured intensity of this line in the \'echelle data is less than 1\% of the intensity of the H$\beta$ line and, therefore, the contribution of $\rm O^{+++}/H$ is very small and does not exceed the accuracy of the total abundance of oxygen (Tables~\ref{tab:Lines} and \ref{t:Chem1}).

\begin{table*}
    \centering{
    \caption{Emission line intensities in the A2 region measured with different types of observation}
    \label{tab:Lines}
    {\small
        \begin{tabular}{l|c|c|c|c} \hline \hline
            \rule{0pt}{10pt}
            & \MC{2}{c}{Long-slit} & \MC{2}{c}{\'Echelle (total flux)}    \\ \hline
            \rule{0pt}{10pt}
            $\lambda_{0}$(\AA) Ion             & $F(\lambda)/F(H\beta)$&$I(\lambda)/I(H\beta)$ & $F(\lambda)/F(H\beta)$&$I(\lambda)/I(H\beta)$  \\ \hline
            3727\,[O\,{\sc ii}]                & 1.276$\pm$0.028 & 1.720$\pm$0.041  &   --                 &   --                  \\
            3869\,[Ne\,{\sc iii}]              & 0.368$\pm$0.009 & 0.475$\pm$0.012  &   0.5763$\pm$0.0453  &   0.6776$\pm$0.0536   \\
            3889\,H\,8                         &       --        &       --         &   0.1877$\pm$0.0194  &   0.2198$\pm$0.0229   \\
            3889\,He\,{\sc i} +\,H8            & 0.149$\pm$0.005 & 0.218$\pm$0.008  &   --                 &   --                  \\
            4026\,He\,{\sc i}                  & 0.009$\pm$0.002 & 0.011$\pm$0.002  &   0.0146$\pm$0.0019  &   0.0167$\pm$0.0022   \\
            4069\,[S\,{\sc ii}]                &       --        &       --         &   0.0106$\pm$0.0016  &   0.0120$\pm$0.0019   \\
            4101\,H$\delta$                    & 0.193$\pm$0.005 & 0.255$\pm$0.007  &   0.2731$\pm$0.0035  &   0.3121$\pm$0.0053   \\
            4340\,H$\gamma$                    & 0.398$\pm$0.009 & 0.467$\pm$0.011  &   0.4843$\pm$0.0144  &   0.5277$\pm$0.0160   \\
            4363\,[O\,{\sc iii}]               & 0.060$\pm$0.003 & 0.067$\pm$0.004  &   0.0693$\pm$0.0010  &   0.0747$\pm$0.0011   \\
            4471\,He\,{\sc i}                  & 0.031$\pm$0.002 & 0.034$\pm$0.002  &   0.0387$\pm$0.0017  &   0.0410$\pm$0.0018   \\
            4658\,[Fe\,{\sc iii}]              &      --         &       --         &   0.0046$\pm$0.0004  &   0.0047$\pm$0.0004   \\
            4686\,He\,{\sc ii}                 &      --         &       --         &   0.0082$\pm$0.0008  &   0.0084$\pm$0.0009   \\
            4713\,He\,{\sc i}                  &                 &                  &   0.0055$\pm$0.0004  &   0.0056$\pm$0.0004   \\
		4740\,[Ar\,{\sc iv}]               & 0.005$\pm$0.002 & 0.005$\pm$0.002  &   0.0036$\pm$0.0004  &   0.0036$\pm$0.0004   \\
            4861\,H$\beta$                     & 1.000$\pm$0.010 & 1.000$\pm$0.011  &   1.0000$\pm$0.0022  &   1.0000$\pm$0.0031   \\
            4922\,He\,{\sc i}                  & 0.015$\pm$0.002 & 0.014$\pm$0.002  &   0.0104$\pm$0.0006  &   0.0103$\pm$0.0006   \\
            4959\,[O\,{\sc iii}]               & 1.847$\pm$0.042 & 1.780$\pm$0.041  &   1.8829$\pm$0.0042  &   1.8515$\pm$0.0041   \\
            5007\,[O\,{\sc iii}]               & 5.588$\pm$0.125 & 5.327$\pm$0.121  &   5.6433$\pm$0.1419  &   5.5112$\pm$0.1395   \\
            5016\,He\,{\sc i}                  &      --         &      --          &   0.0204$\pm$0.0009  &   0.0199$\pm$0.0009   \\
            5192\,[Ar\,{\sc iii}]              &      --         &      --          &   0.0021$\pm$0.0002  &   0.0020$\pm$0.0002   \\
            5199\,[N\,{\sc i}]                 &      --         &      --          &   0.0043$\pm$0.0003  &   0.0041$\pm$0.0003   \\
            5200\,[N\,{\sc i}]                 &      --         &      --          &   0.0012$\pm$0.0003  &   0.0011$\pm$0.0002   \\
            5518\,[Cl\,{\sc iii}]              & 0.004$\pm$0.001 & 0.004$\pm$0.001  &   0.0068$\pm$0.0009  &   0.0062$\pm$0.0008   \\
            5538\,[Cl\,{\sc iii}]              & 0.003$\pm$0.001 & 0.002$\pm$0.001  &   0.0032$\pm$0.0005  &   0.0029$\pm$0.0004   \\
            5755\,[N\,{\sc ii}]                &      --         &       --         &   0.0015$\pm$0.0002  &   0.0013$\pm$0.0002   \\
            5876\,He\,{\sc i}                  & 0.145$\pm$0.004 & 0.115$\pm$0.003  &   0.1301$\pm$0.0012  &   0.1136$\pm$0.0010   \\
            6300\,[O\,{\sc i}]                 & 0.034$\pm$0.002 & 0.025$\pm$0.001  &   0.0299$\pm$0.0009  &   0.0249$\pm$0.0008   \\
            6312\,[S\,{\sc iii}]               & 0.023$\pm$0.001 & 0.017$\pm$0.001  &   0.0232$\pm$0.0004  &   0.0193$\pm$0.0003   \\
            6364\,[O\,{\sc i}]                 & 0.012$\pm$0.001 & 0.009$\pm$0.001  &   0.0096$\pm$0.0005  &   0.0079$\pm$0.0004   \\
            6548\,[N\,{\sc ii}]                & 0.039$\pm$0.001 & 0.028$\pm$0.001  &   0.0266$\pm$0.0004  &   0.0216$\pm$0.0003   \\
            6563\,H$\alpha$                    & 4.020$\pm$0.088 & 2.840$\pm$0.069  &   3.4883$\pm$0.0067  &   2.8310$\pm$0.0060   \\
            6583\,[N\,{\sc ii}]                & 0.118$\pm$0.006 & 0.083$\pm$0.004  &   0.0814$\pm$0.0008  &   0.0659$\pm$0.0006   \\
            6678\,He\,{\sc i}                  & 0.043$\pm$0.002 & 0.030$\pm$0.001  &   0.0372$\pm$0.0011  &   0.0299$\pm$0.0009   \\
            6716\,[S\,{\sc ii}]                & 0.186$\pm$0.004 & 0.128$\pm$0.003  &   0.1500$\pm$0.0007  &   0.1199$\pm$0.0005   \\
            6731\,[S\,{\sc ii}]                & 0.137$\pm$0.003 & 0.094$\pm$0.003  &   0.1146$\pm$0.0005  &   0.0915$\pm$0.0004   \\
            7065\,He\,{\sc i}                  &      --         &      --          &   0.0334$\pm$0.0011  &   0.0259$\pm$0.0008   \\
            7136\,[Ar\,{\sc iii}]              &      --         &      --          &   0.1129$\pm$0.0009  &   0.0869$\pm$0.0007   \\
            7281\,He\,{\sc i}                  &      --         &      --          &   0.0077$\pm$0.0004  &   0.0059$\pm$0.0003   \\
            7320\,[O\,{\sc ii}]                &      --         &      --          &   0.0341$\pm$0.0010  &   0.0258$\pm$0.0008   \\
            7330\,[O\,{\sc ii}]                &      --         &      --          &   0.0285$\pm$0.0007  &   0.0216$\pm$0.0006   \\
            7751\,[Ar\,{\sc iii}]              &      --         &      --          &   0.0284$\pm$0.0009  &   0.0208$\pm$0.0007   \\
            8374\,P\,21                        &      --         &      --          &   0.0027$\pm$0.0003  &   0.0019$\pm$0.0002   \\
            8392\,P\,20                        &      --         &      --          &   0.0038$\pm$0.0003  &   0.0026$\pm$0.0002   \\
            8413\,P\,19                        &      --         &      --          &   0.0044$\pm$0.0003  &   0.0031$\pm$0.0002   \\
            8438\,P\,18                        &      --         &      --          &   0.0048$\pm$0.0003  &   0.0034$\pm$0.0002   \\
            8467\,P\,17                        &      --         &      --          &   0.0052$\pm$0.0003  &   0.0036$\pm$0.0002  \\
            8502\,P\,16                        &      --         &      --          &   0.0065$\pm$0.0004  &   0.0045$\pm$0.0002  \\
            8545\,P\,15                        &      --         &      --          &   0.0077$\pm$0.0005  &   0.0053$\pm$0.0003  \\
            8598\,P\,14                        &      --         &      --          &   0.0097$\pm$0.0004  &   0.0067$\pm$0.0003  \\
            8665\,P\,13                        &      --         &      --          &   0.0110$\pm$0.0006  &   0.0076$\pm$0.0004  \\
            \hline
            EW(abs)\ \AA\                      & \MC {2}{c}{2.40$\pm$0.25}          & \MC {2}{c}{2.00$\pm$0.24} \\
            EW(H$\beta$)\ \AA\                 & \MC {2}{c}{ 137$\pm$6}             & \MC {2}{c}{ 114$\pm$4}     \\
            C(H$\beta$)\ dex                   & \MC {2}{c}{0.36$\pm$0.050}         & \MC {2}{c}{0.27$\pm$0.01} \\
            E(B-V)\ mag                        & \MC {2}{c}{0.25$\pm$0.040}         & \MC {2}{c}{0.18$\pm$0.01} \\
            \hline
        \end{tabular}
    }}
\end{table*}

\begin{table*}
    \centering{
        \caption{Physical conditions and chemical element abundances in the A2 region for different types of observation
                 and in the different kinematic subsystems found with the \'echelle data}
        \label{t:Chem1}
        \begin{tabular}{l|l|l|l|l|l} \hline
\MC{1}{c|}{Value}    & \MC{1}{c|}{Long-slit}& \MC{2}{c|}{\'Echelle (total flux)}  & \MC{2}{c}{\'Echelle (subsystems)}     \\
                     &                      & $T_{\rm e}$(OII) &  $T_{\rm e}$(NII)&\MC{1}{c}{Narrow}   &  \MC{1}{c}{Wide} \\ 
\MC{1}{c|}{(1)}      &\MC{1}{c|}{(2)}       &  \MC{1}{c|}{(3)} &  \MC{1}{c|}{(4)} &  \MC{1}{c|}{(5)}   &  \MC{1}{c}{(6)} \\  \hline
$T_{\rm e}$(OIII)(K)\                & 12,551$\pm$278    & 12,929$\pm$121      &  12,929$\pm$121    & 13,194$\pm$248    &  12,169$\pm$494   \\
$T_{\rm e}$(OII)(K)\                 & 12,295$\pm$224    &{\bf 12,591$\pm$92}  &  12,591$\pm$92     & 12,788$\pm$180    &  13,160$\pm$380   \\
$T_{\rm e}$(NII)(K)\                 & --                & 11,790$\pm$924      &{\bf 11,790$\pm$924}& --                &   --              \\
$T_{\rm e}$(SIII)(K)\                & 11,836$\pm$354    & 12,310$\pm$149      &  12,310$\pm$149    & 12,633$\pm$298    &  12,784$\pm$813   \\
$N_{\rm e}$(SII)(cm$^{-3}$)\         &  55$\pm$52        & 107$\pm$14          &  108$\pm$13        & 111$\pm$44        &  88$\pm$73        \\
                                     &                   &                     &                    &                   &                   \\
O$^{+}$/H$^{+}$($\times$10$^5$)\     & 3.054$\pm$0.216   & 2.581$\pm$0.128     &  3.626$\pm$1.674   & 2.790$\pm$0.310   &  3.584$\pm$0.766  \\
O$^{++}$/H$^{+}$($\times$10$^5$)\    & 9.476$\pm$0.629   & 9.020$\pm$0.293     &  9.020$\pm$0.293   & 8.160$\pm$0.469   &  11.700$\pm$1.439 \\
O$^{+++}$/H$^{+}$($\times$10$^5$)\   & --                & 0.103$\pm$0.011     &  0.112$\pm$0.019   & --                &  0.807$\pm$0.134  \\
O/H($\times$10$^5$)\                 & 12.530$\pm$0.665  & 11.700$\pm$0.320    &  12.760$\pm$1.700  & 10.910$\pm$0.534  &  16.090$\pm$0.134 \\
12+log(O/H)\                         & 8.10$\pm$0.02     & 8.07$\pm$0.01       &  8.11$\pm$0.06     & 8.04$\pm$0.02     &  8.21$\pm$0.04    \\
                                     &                   &                     &                    &                   &                   \\
N$^{+}$/H$^{+}$($\times$10$^7$)\     & 9.551$\pm$0.551   & 7.167$\pm$0.131     &  8.332$\pm$1.535   & 6.676$\pm$0.251   &  8.492$\pm$0.666  \\
ICF(N)\                              & 4.011             & 4.397               &  3.480             & 3.931             &  1.248            \\
N/H($\times$10$^5$)\                 & 0.38$\pm$0.02     & 0.32$\pm$0.01       &  0.29$\pm$0.05     & 0.26$\pm$0.01     &  0.38$\pm$0.03    \\
12+log(N/H)\                         & 6.58$\pm$0.03     & 6.50$\pm$0.01       &  6.46$\pm$0.08     & 6.41$\pm$0.02     &  6.58$\pm$0.03    \\
log(N/O)\                            & -1.51$\pm$0.03    & -1.57$\pm$0.01      &  -1.64$\pm$0.10    & -1.62$\pm$0.03    &  -1.62$\pm$0.06   \\
                                     &                   &                     &                    &                   &                   \\
Ne$^{++}$/H$^{+}$($\times$10$^5$)\   & 2.224$\pm$0.176   & 2.876$\pm$0.244     &  2.876$\pm$0.244   & 2.731$\pm$0.286   &  --               \\
ICF(Ne)\                             & 1.103             & 1.097               &  1.125             & 1.107             &  --               \\
Ne/H($\times$10$^5$)\                & 2.46$\pm$0.19     & 3.16$\pm$0.27       &  3.24$\pm$0.27     & 3.03$\pm$0.32     &  --               \\
12+log(Ne/H)\                        & 7.39$\pm$0.03     & 7.50$\pm$0.04       &  7.51$\pm$0.04     & 7.48$\pm$0.05     &  --               \\
log(Ne/O)\                           & -0.71$\pm$0.04    & -0.57$\pm$0.04      &  -0.60$\pm$0.07    & -0.56$\pm$0.05    &  --               \\
                                     &                   &                     &                    &                   &                   \\
S$^{+}$/H$^{+}$($\times$10$^7$)\     & 3.145$\pm$0.134   & 9.900$\pm$2.816     &  9.921$\pm$2.823   & 6.540$\pm$1.928   &  3.702$\pm$0.222  \\
S$^{++}$/H$^{+}$($\times$10$^7$)\    & 19.790$\pm$2.515  & 19.250$\pm$0.900    &  19.250$\pm$0.900  & 17.190$\pm$1.608  &  20.130$\pm$4.913 \\
ICF(S)\                              & 1.141             & 1.164               &  1.111             & 1.134             &  1.337            \\
S/H($\times$10$^7$)\                 & 26.18$\pm$2.87    & 33.94$\pm$3.44      &  32.42$\pm$3.29    & 26.91$\pm$2.85    &  31.85$\pm$6.57   \\
12+log(S/H)\                         & 6.42$\pm$0.05     & 6.53$\pm$0.04       &  6.51$\pm$0.04     & 6.43$\pm$0.05     &  6.50$\pm$0.09    \\
log(S/O)\                            & -1.68$\pm$0.05    & -1.54$\pm$0.05      &  -1.59$\pm$0.07    & -1.61$\pm$0.05    &  -1.70$\pm$0.10   \\
                                     &                   &                     &                    &                   &                   \\
Ar$^{++}$/H$^{+}$($\times$10$^7$)\   & --                & 5.055$\pm$0.130     &  5.055$\pm$0.130   & 4.694$\pm$0.280   &  5.827$\pm$0.788  \\
Ar$^{+++}$/H$^{+}$($\times$10$^7$)\  & --                & 0.557$\pm$0.060     &  0.557$\pm$0.060   & 0.613$\pm$0.066   &  0.000$\pm$0.000  \\
ICF(Ar)\                             & --                & 1.008               &  1.013             & 1.010             &  1.067            \\
Ar/H($\times$10$^7$)\                & --                & 5.65$\pm$0.14       &  5.69$\pm$0.15     & 5.36$\pm$0.29     &  6.22$\pm$0.84    \\
12+log(Ar/H)\                        & --                & 5.75$\pm$0.01       &  5.75$\pm$0.01     & 5.73$\pm$0.02     &  5.79$\pm$0.06    \\
log(Ar/O)\                           & --                & -2.32$\pm$0.02      &  -2.35$\pm$0.06    & -2.31$\pm$0.03    &  -2.41$\pm$0.07   \\
                                     &                   &                     &                    &                   &                   \\
Cl$^{++}$/H$^{+}$($\times$10$^7$)\   & 0.217$\pm$0.055   & 0.315$\pm$0.034     &  0.315$\pm$0.034   & 0.345$\pm$0.041   &  --               \\
ICF(Cl)\                             & 1.356             & 1.393               &  1.312             & 1.345             &  --               \\
Cl/H($\times$10$^7$)\                & 0.29$\pm$0.07     & 0.44$\pm$0.05       &  0.41$\pm$0.04     & 0.47$\pm$0.06     &  --               \\
12+log(Cl/H)\                        & 4.47$\pm$0.11     & 4.64$\pm$0.05       &  4.62$\pm$0.05     & 4.67$\pm$0.05     &  --               \\
log(Cl/O)\                           & -3.63$\pm$0.11    & -3.43$\pm$0.05      &  -3.49$\pm$0.07    & -3.37$\pm$0.06    &  --               \\
                                     &                   &                     &                    &                   &                   \\
Fe$^{++}$/H$^{+}$($\times$10$^7$)\   & --                & 1.425$\pm$0.122     &  1.728$\pm$0.433   & 1.590$\pm$0.141   &  --               \\
ICF(Fe)\                             & --                & 6.006               &  4.636             & 5.304             &  --               \\
log(Fe/O)\                           & --                & -2.14$\pm$0.04      &  -2.20$\pm$0.12    & -2.12$\pm$0.04    &  --               \\
$[$O/Fe$]$\                          & --                & 0.92$\pm$0.04       &  0.98$\pm$0.12     & 0.90$\pm$0.04     &  --               \\
\hline
        \end{tabular}
    }
\end{table*}

\section{Results}
\label{txt:results}

Intensities of measured emission lines $\rm F(\lambda)$ reduced to intensities
of the H$\beta$ line, and the same intensities $\rm I(\lambda)$ corrected
for interstellar absorption and underlying Balmer absorption are presented
in Table~\ref{tab:Lines} for both spectra, 
together with the equivalent width $EW$(H$\beta$) of the H$\beta$ emission line,
the equivalent width $EW$(abs) of the Balmer absorption lines, 
the extinction coefficient $C$(H$\beta$) and the extinction E(B-V) value calculated from it.
The obtained extinction is the total extinction on the line-of-sight, being the 
sum of (1) the intrinsic extinction in the \IC\ galaxy, 
(2) the intergalactic extinction, and 
(3) the foreground extinction in our Galaxy.
The derived electron temperatures T$_{\rm e}$ and electron densities N$_{\rm e}$, and
the calculated abundances of oxygen, nitrogen, neon, sulphur, argon, chlorine and iron are given
in columns (2)--(4) of Table~\ref{t:Chem1}.

\subsection{Physical parameters and abundances of chemical elements in 
            the long-slit and \'echelle observations}
\label{txt:comp}

The line intensities presented in Table~\ref{tab:Lines} show that
two quite close but not exactly overlapping
parts of the \ion{H}{ii} region A2 were apparently observed. 
This is obvious since in the case of the long-slit observation,
a slit 1\farcs5 wide was used, while in the case of the \'echelle observations 
a circular fibre of 2\farcs23 in diameter was used.
The image quality in both observations was very close to 1\farcs4 and 1\farcs5 respectively.
Nevertheless, it should be noted that the intensities of most of the lines
correlate quite well with each other.
Exactly the same conclusion can be drawn regarding the measured equivalent width
of the emission line H$\beta$ and the  estimated equivalent width 
of the Balmer absorption lines $EW$(abs),
which are in good agreement with each other taking into account their errors.
This agreement can be seen in the calculated temperatures, densities and chemical element abundances
given in columns (2)--(4) of Table~\ref{t:Chem1}.
The electron densities calculated from the
sulphur lines, [\ion{S}{ii}]~$\lambda\lambda$6717,6731~\AA,
differ most strongly
and the main reason seems to be that the line [\ion{S}{ii}]~$\lambda$6731~\AA\
is at the very edge of the long-slit spectrum (Figure~\ref{fig:IC4662_A2_long}).
This could introduce an additional systematic error into its flux measurement.
However, the measured electron densities are so small in absolute
value (and this is normal in \ion{H}{ii} regions) that even a difference of a factor of two
does not result in a large difference in calculated abundances
between the two spectra.
As already mentioned in Section~\ref{txt:analysis}
in the case of the \'echelle spectrum, the abundance of $O^{+++}$ was also taken into account
for the calculation of the total abundance of oxygen.
This was calculated using the intensity of the \ion{He}{ii}~$\lambda$4686~\AA\ line
\citet{2008MNRAS.388.1667K}. 
However, as can be seen in Table~\ref{t:Chem1}, 
the abundance of this ion is less than 0.5\% and therefore its contribution
is negligibly small.

Overall, with an exposure difference of a factor of two, 
the \'echelle spectrum allows measurement of 
emission lines at 0.12--0.15\% of H$\beta$ intensity with a
signal-to-noise ratio about four, while in the long-slit spectrum the weakest lines
are lines at 0.3\% of the H$\beta$ intensity with a signal-to-noise ratio about three.
Of course, this is due to the difference in spectral resolution and it is quite obvious
that this advantage of \'echelle spectroscopy will disappear quickly towards
fainter objects.

To estimate the possible systematic error associated with different cold zone temperature estimates using the approximated $T_{\rm e}$(OII) temperature \citep{2006A&A...448..955I} for the long-slit spectrum and a direct temperature calculation $T_{\rm e}$(NII) based on line ratios [\ion{N}{ii}]~$\lambda\lambda$5755,6548,6584~\AA, column (3) of Table~\ref{t:Chem1} gives calculated abundances using the approximated $T_{\rm e}$(OII), and in column (4) the calculated abundances using directly calculated temperature $T_{\rm e}$(NII). { The specific value of $T_{\rm e}$(OII) used in each case is highlighted in bold font in each column.} It can be concluded that although, in general, elemental abundances using the directly computed temperature $T_{\rm e}$(NII) are higher, the two values are consistent to within the cited errors. Unfortunately, the intensity of the auroral line [\ion{N}{ii}]~ $\lambda$5755~\AA\ is very small (0.15\% of the intensity of the line H$\beta$) and therefore the signal-to-noise ratio for it is only about seven.

The comparison of abundances calculated using the same methodology
for two different observational methods and presented in Table~\ref{t:Chem1} (columns (2) and (3)),
allows us to conclude that the calculated abundances agree to within the cited uncertainties.
For that reason, the HRS data can be used to study the abundances of chemical
elements in bright \ion{H}{ii} regions and planetary nebulae.

\subsection{Comparison with previous results}
\label{txt:phys}

The measured extinction coefficient, $C$(H$\beta$), as well as the extinction value
$E(B-V) = 0.68 \cdot \mathrm{C(H\beta)}$, in region A2
also agree well for both types of observation with an accuracy of 1.1$\sigma$.
If we take the foreground extinction of our Galaxy in the \IC\ direction
as 0.19--0.23~mag \citep{1998ApJ...500..525S,2011ApJ...737..103S},
then there is almost no additional extinction in the line-of-sight outside the Milky Way.
Here it should be noted that previous work to estimate extinction
used either the Balmer line ratio H$\gamma$/H$\beta$ \citep{2001A&A...369..797H}
or H$\delta$, H$\gamma$ and H$\beta$ \citep{2009A&A...499..455C}.
Where only the H$\delta$, H$\gamma$ and H$\beta$ lines are used
to determine the extinction, the total measured value $E(B-V)=0.36$~mag immediately drops to
$E(B-V)=0.16$~mag, which is in perfect agreement with 
the measured average extinction value of 0.162~mag for region A2 from \citet{2009A&A...499..455C}.

Our derived abundances of oxygen and neon in the A2 region coincide
within the cited uncertainties with previous estimates of abundances of these 
elements from \citet{1990A&A...234...99H,2001A&A...369..797H,2009A&A...499..455C}.
The nitrogen abundance for the region A2 was obtained earlier in \citet{2001A&A...369..797H}
and matches very well with the nitrogen abundance calculated from our data.
\citet{2009Ap&SS.324..147G} estimated neon and sulphur content
using data from Spitzer/IRS infrared spectroscopy.
The abundances of these elements are slightly lower than our values,
but, unfortunately, there are no errors given in that paper.
If we assume that their measured abundances have errors equal to ours ($\sigma_{el}$, where subscript 'el' refers to the particular element under discussion),
then the difference does not exceed 2 to 3$\sigma_{el}$.
Also, \citet{2009Ap&SS.324..147G} point out a possible
systematic shift in their abundances due to the broadening of the measured line,
Hu$\alpha$, of the Humphrey series, since the intensities of all
lines used by them were related to the intensity of this line.

The abundances of argon, chlorine and iron in the A2 region have been determined 
for the first time in our work.
It is known that $\alpha$-elements determined from the emission spectra of \ion{H}{ii} regions
are formed in massive stars. As a consequence, it is logical to assume that the abundances of these
elements in \ion{H}{ii} regions are proportional to each other,
and for that reason their ratios log(Ne/O), log(S/O), log(Ar/O) and log(Cl/O) should be constant.
For example, the results of \citet{2006A&A...448..955I}
based on a very large sample of \ion{H}{ii} galaxies from the DR3 release
of the SDSS survey \citep{2005AJ....129.1755A}
show that this assumption is correct.
As can be seen by comparing the values given in our Table~\ref{t:Chem1}
and the dependencies given in \citet{2006A&A...448..955I},
all the abundances of $\alpha$-elements determined by us, 
as well as their relations to the oxygen abundance,
agree very well with these dependencies, which is an additional confirmation of
the correctness of the abundances reported here.
The iron abundance determined with the weak line [\ion{Fe}{iii}]~$\lambda$4658~\AA\
is also consistent with the relationship found in \citet{2006A&A...448..955I}
and most likely reflects the fact of iron deposition on dust particles.

\begin{figure}
    \includegraphics[clip=,angle=0,width=0.48\textwidth]{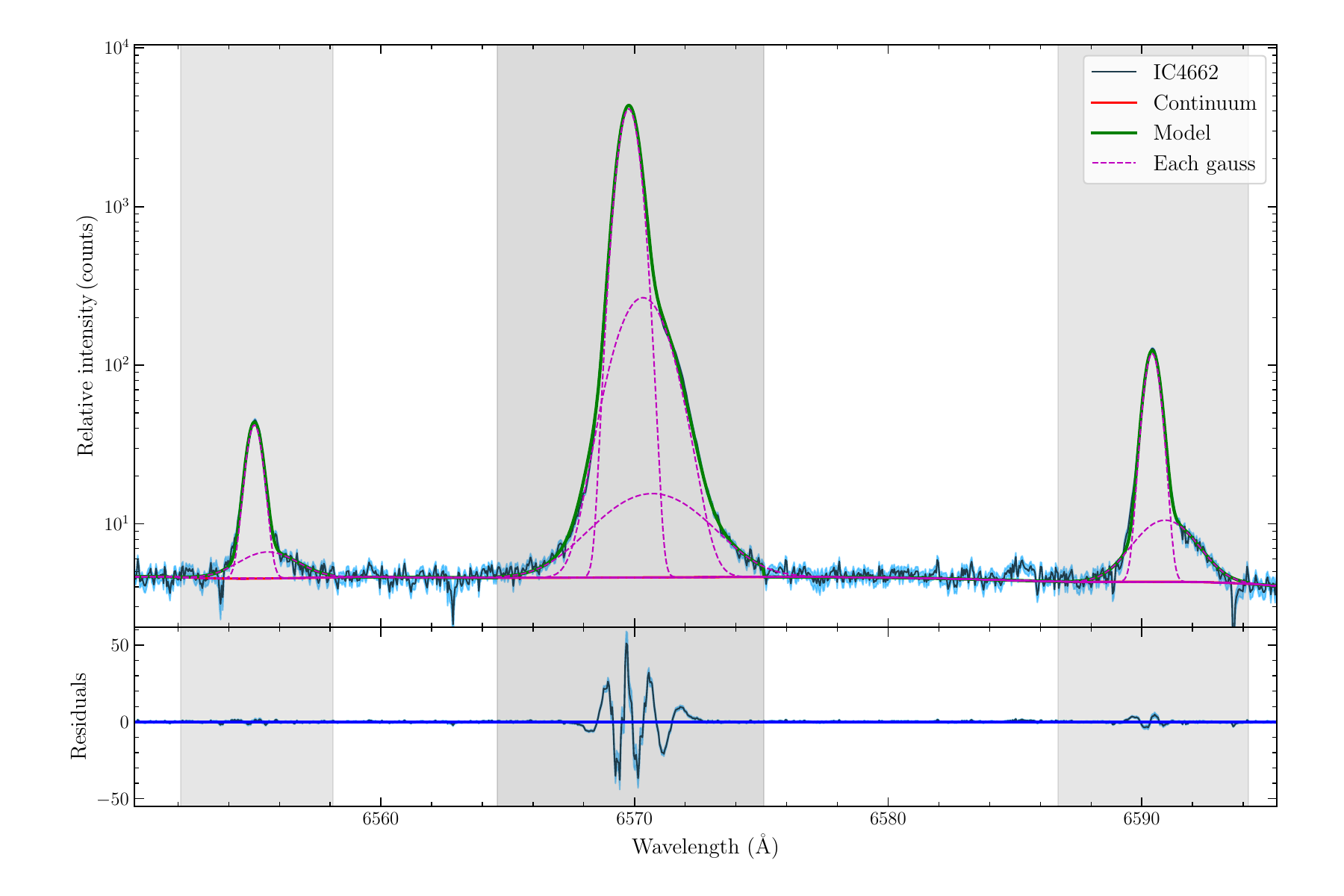}
    \includegraphics[clip=,angle=0,width=0.48\textwidth]{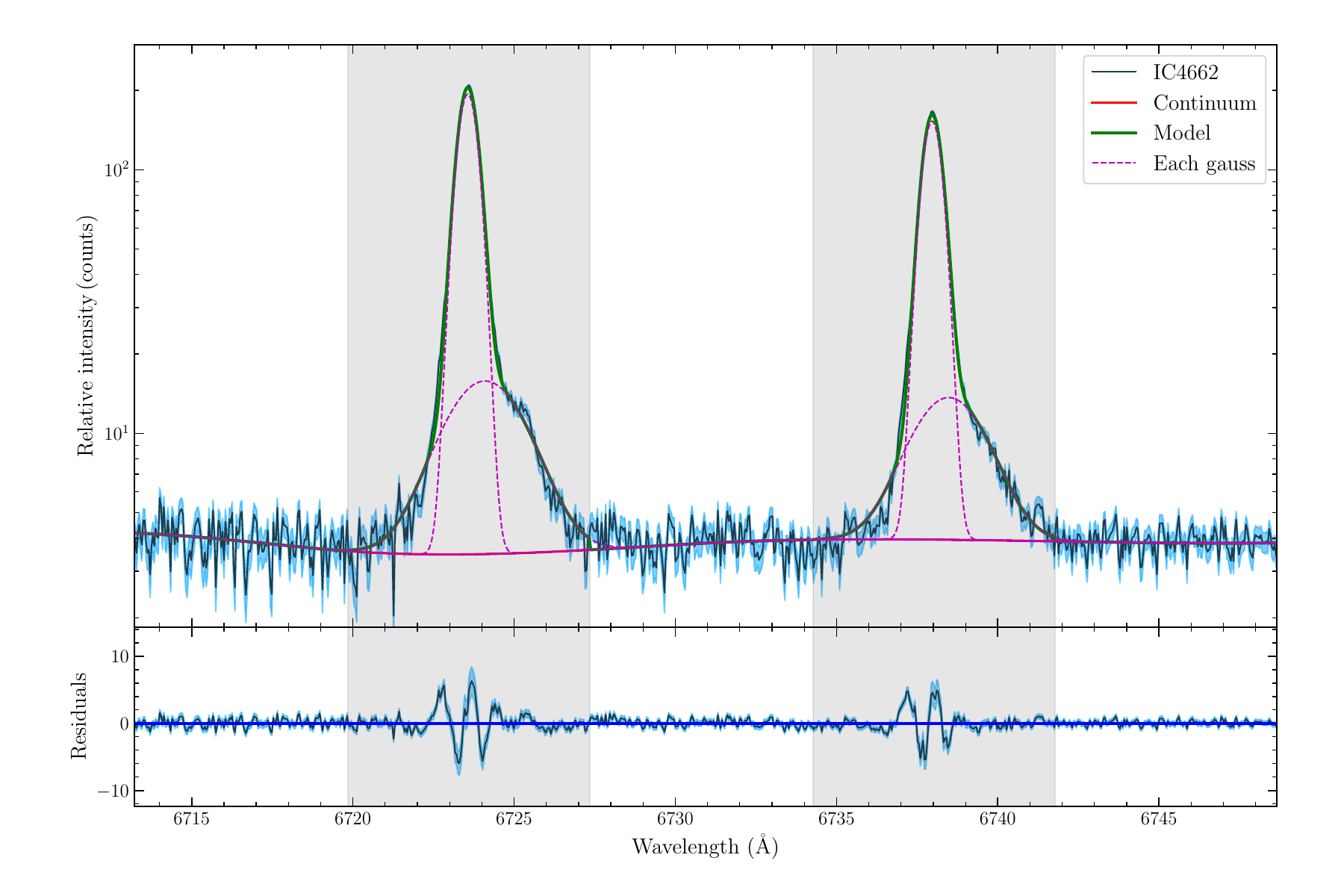}
    \caption{%
        Components of the different emission lines found in the analysis of the \'echelle spectrum.
        The spectral region of the H$\alpha$ line and nitrogen lines [\ion{N}{ii}]~$\lambda\lambda$6548,6583~\AA\
        is shown in the left plot and the spectral region of the sulphur doublet [\ion{S}{ii}]~$\lambda\lambda$6716,6731~\AA\
        in the right plot. Both plots consist of two panels.
        In the upper panels the observed spectrum, the constructed continuum, 
        the emission line components and the full model profile of each line 
        (almost identical to the observed one) are shown on a logarithmic scale.
        The bottom panels show the difference between the observed spectrum and the model spectrum.
        The vertical grey bars show the regions for each line within which the
        approximation by Gaussians was done.
        \label{fig:dual}}
\end{figure}

\begin{table*}
    \centering{
        \caption{Intensities of the narrow and wide components of emission lines}
        \label{tab:Lines_kin}
        \begin{tabular}{l|cccc|cccc} \hline
            & \MC{4}{c|}{Narrow component A2} & \MC{4}{c}{Wide component A2}    \\ \hline
$\lambda_{0}$(\AA) Ion & F($\lambda$)/F(H$\beta$)&I($\lambda$)/I(H$\beta$)&V$_{hel}$    &$\sigma_{gas}$ &F($\lambda$)/F(H$\beta$)&I($\lambda$)/I(H$\beta$) & V$_{hel}$      & $\sigma_{gas}$  \\ 
            &                         &                        &(km s$^{-1}$)&(km s$^{-1}$)  &                        &                         &  (km s$^{-1}$) &  (km s$^{-1}$)  \\ \hline
3869\,[Ne\,{\sc iii}]  &   0.5865$\pm$0.0492     &   0.6879$\pm$0.0586    &  299.51     &12.42$\pm$0.66 &   0.500$\pm$0.243      &   0.567$\pm$0.276       &     320.86     &  32.68$\pm$3.41 \\
4026\,He\,{\sc i}      &   0.0171$\pm$0.0023     &   0.0194$\pm$0.0026    &  300.30     & 8.49$\pm$1.15 &         ---            &         ---             &     ---        &        ---      \\
4069\,[S\,{\sc ii}]    &   0.0123$\pm$0.0019     &   0.0140$\pm$0.0022    &  300.96     &11.60$\pm$2.02 &         ---            &         ---             &     ---        &        ---      \\
4102\,H$\delta$        &   0.2770$\pm$0.0106     &   0.3150$\pm$0.0126    &  299.77     &10.74$\pm$0.15 &   0.250$\pm$0.024      &   0.274$\pm$0.033       &     325.39     &  36.04$\pm$2.94 \\
4340\,H$\gamma$        &   0.4890$\pm$0.0167     &   0.5315$\pm$0.0187    &  299.70     &11.42$\pm$0.06 &   0.456$\pm$0.028      &   0.485$\pm$0.034       &     328.40     &  41.57$\pm$1.20 \\
4363\,[O\,{\sc iii}]   &   0.0699$\pm$0.0028     &   0.0753$\pm$0.0030    &  300.65     &10.47$\pm$0.16 &   0.065$\pm$0.007      &   0.069$\pm$0.007       &     320.23     &  34.11$\pm$2.51 \\
4471\,He\,{\sc i}      &   0.0452$\pm$0.0020     &   0.0478$\pm$0.0021    &  301.84     &10.58$\pm$0.28 &         ---            &         ---             &     ---        &        ---      \\
4658\,[Fe\,{\sc iii}]  &   0.0053$\pm$0.0004     &   0.0055$\pm$0.0005    &  305.59     &15.13$\pm$0.82 &         ---            &         ---             &     ---        &        ---      \\
4686\,He\,{\sc ii}     &        ---              &        ---             &  ---        &    ---        &   0.058$\pm$0.006      &   0.059$\pm$0.006       &     321.48     &  19.23$\pm$3.23 \\
4861\,H$\beta$         &   1.0000$\pm$0.0018     &   1.0000$\pm$0.0024    &  299.72     &10.98$\pm$0.02 &   1.000$\pm$0.012      &   1.000$\pm$0.016       &     325.75     &  44.59$\pm$0.41 \\
4959\,[O\,{\sc iii}]   &   1.7951$\pm$0.0556     &   1.7667$\pm$0.0551    &  301.10     &10.82$\pm$0.04 &   2.084$\pm$0.075      &   2.062$\pm$0.075       &     328.75     &  37.32$\pm$1.22 \\
5007\,[O\,{\sc iii}]   &   5.4047$\pm$0.1653     &   5.2833$\pm$0.1627    &  301.20     &10.79$\pm$0.03 &   6.052$\pm$0.206      &   5.956$\pm$0.204       &     330.50     &  35.77$\pm$1.00 \\
5538\,[Cl\,{\sc iii}]  &   0.0037$\pm$0.0006     &   0.0034$\pm$0.0005    &  298.27     &11.39$\pm$2.73 &         ---            &         ---             &     ---        &        ---      \\
5876\,He\,{\sc i}      &   0.1307$\pm$0.0045     &   0.1145$\pm$0.0041    &  302.23     &11.90$\pm$0.10 &   0.127$\pm$0.008      &   0.115$\pm$0.008       &     330.88     &  38.50$\pm$1.66 \\
6300\,[O\,{\sc i}]     &   0.0289$\pm$0.0011     &   0.0241$\pm$0.0010    &  297.13     &11.26$\pm$0.14 &   0.036$\pm$0.003      &   0.032$\pm$0.003       &     320.75     &  46.61$\pm$2.60 \\
6312\,[S\,{\sc iii}]   &   0.0227$\pm$0.0009     &   0.0189$\pm$0.0008    &  298.93     &11.84$\pm$0.19 &   0.027$\pm$0.003      &   0.023$\pm$0.002       &     321.54     &  38.66$\pm$5.04 \\
6364\,[O\,{\sc i}]     &   0.0112$\pm$0.0005     &   0.0093$\pm$0.0005    &  298.94     &13.88$\pm$0.25 &         ---            &         ---             &     ---        &        ---      \\
6548\,[N\,{\sc ii}]    &   0.0254$\pm$0.0010     &   0.0207$\pm$0.0008    &  297.81     &10.84$\pm$0.09 &   0.034$\pm$0.003      &   0.029$\pm$0.003       &     317.69     &  47.31$\pm$2.59 \\
6563\,H$\alpha$        &   3.4681$\pm$0.1068     &   2.8269$\pm$0.0951    &  299.75     &10.65$\pm$0.01 &   3.327$\pm$0.122      &   2.844$\pm$0.113       &     327.41     &  40.38$\pm$0.39 \\
6583\,[N\,{\sc ii}]    &   0.0783$\pm$0.0027     &   0.0637$\pm$0.0023    &  300.88     &11.10$\pm$0.05 &   0.100$\pm$0.005      &   0.085$\pm$0.005       &     318.16     &  45.93$\pm$1.19 \\
6678\,He\,{\sc i}      &   0.0365$\pm$0.0013     &   0.0294$\pm$0.0011    &  302.47     &10.52$\pm$0.11 &   0.042$\pm$0.003      &   0.035$\pm$0.003       &     321.51     &  40.74$\pm$2.61 \\
6716\,[S\,{\sc ii}]    &   0.1414$\pm$0.0005     &   0.1135$\pm$0.0016    &  301.34     &11.56$\pm$0.06 &   0.202$\pm$0.004      &   0.171$\pm$0.004       &     318.15     &  48.92$\pm$0.94 \\
6731\,[S\,{\sc ii}]    &   0.1083$\pm$0.0004     &   0.0869$\pm$0.0013    &  300.83     &11.40$\pm$0.07 &   0.152$\pm$0.003      &   0.129$\pm$0.003       &     317.00     &  49.01$\pm$1.28 \\
7065\,He\,{\sc i}      &   0.0349$\pm$0.0012     &   0.0272$\pm$0.0011    &  296.94     &11.21$\pm$0.06 &   0.024$\pm$0.002      &   0.020$\pm$0.002       &     329.19     &  40.32$\pm$2.51 \\
7136\,[Ar\,{\sc iii}]  &   0.1098$\pm$0.0036     &   0.0850$\pm$0.0032    &  299.17     &10.23$\pm$0.05 &   0.132$\pm$0.006      &   0.108$\pm$0.006       &     317.33     &  41.45$\pm$1.18 \\
7320\,[O\,{\sc ii}]    &   0.0398$\pm$0.0006     &   0.0303$\pm$0.0007    &  301.73     &12.34$\pm$1.40 &   0.054$\pm$0.002      &   0.044$\pm$0.002       &     319.78     &  49.45$\pm$4.83 \\
7330\,[O\,{\sc ii}]    &   0.0332$\pm$0.0008     &   0.0253$\pm$0.0008    &  300.51     &12.91$\pm$1.96 &   0.045$\pm$0.002      &   0.037$\pm$0.002       &     321.85     &  48.16$\pm$3.15 \\
7751\,[Ar\,{\sc iii}]  &   0.0283$\pm$0.0010     &   0.0209$\pm$0.0009    &  299.86     & 9.88$\pm$0.07 &   0.029$\pm$0.002      &   0.023$\pm$0.002       &     322.76     &  47.25$\pm$2.43 \\
8413\,P\,19            &   0.0052$\pm$0.0004     &   0.0036$\pm$0.0003    &  301.44     &12.68$\pm$0.45 &         ---            &         ---             &     ---        &        ---      \\
8438\,P\,18            &   0.0056$\pm$0.0003     &   0.0040$\pm$0.0003    &  300.45     &11.70$\pm$0.25 &         ---            &         ---             &     ---        &        ---      \\
8467\,P\,17            &   0.0061$\pm$0.0003     &   0.0043$\pm$0.0003    &  300.81     &11.52$\pm$0.21 &         ---            &         ---             &     ---        &        ---      \\
8502\,P\,16            &   0.0076$\pm$0.0004     &   0.0053$\pm$0.0003    &  300.86     &12.45$\pm$0.24 &         ---            &         ---             &     ---        &        ---      \\
8545\,P\,15            &   0.0090$\pm$0.0005     &   0.0063$\pm$0.0004    &  300.20     &12.64$\pm$0.33 &         ---            &         ---             &     ---        &        ---      \\
8598\,P\,14            &   0.0114$\pm$0.0005     &   0.0079$\pm$0.0004    &  300.54     &11.78$\pm$0.16 &         ---            &         ---             &     ---        &        ---      \\
8665\,P\,13            &   0.0128$\pm$0.0006     &   0.0089$\pm$0.0005    &  302.34     &11.58$\pm$0.19 &         ---            &         ---             &     ---        &        ---      \\
& & \\
C(H$\beta$)\ dex               & \MC {2}{c}{0.27$\pm$0.04}    &&          & \MC {2}{c}{0.23$\pm$0.05} \\
EW(H$\beta$)\ \AA\             & \MC {2}{c}{  98$\pm$ 2}      &&          & \MC {2}{c}{  16$\pm$ 1}   \\
            \hline
        \end{tabular}
    }
\end{table*}

\begin{table}
    \centering{
        \caption{Intensities of the very wide components}
        \label{tab:Lines_kin1}
        \begin{tabular}{l|ccc} \hline
            & \MC{3}{c}{Very Wide Component}    \\ \hline
            $\lambda_{0}$(\AA) Ion    & F($\lambda$)/F(H$\beta$)&V$_{hel}$    &$\sigma_{gas}$ \\ 
            &                         &(km s$^{-1}$)&(km s$^{-1}$)  \\ \hline                    
            4686\,He\,{\sc ii}        &    44.64$\pm$2.38       &  332.78     &393.1$\pm$25.1 \\
            4959\,[O\,{\sc iii}]      &   106.01$\pm$3.46       &  349.46     & 55.8$\pm$3.2  \\
            5007\,[O\,{\sc iii}]      &   342.37$\pm$8.01       &  348.26     & 60.4$\pm$1.5  \\
            6563\,H$\alpha$           &    69.27$\pm$6.76       &  344.83     &109.5$\pm$8.9  \\ \hline
        \end{tabular}
    }
\end{table}

\begin{figure}
    \includegraphics[clip=,angle=0,width=0.48\textwidth]{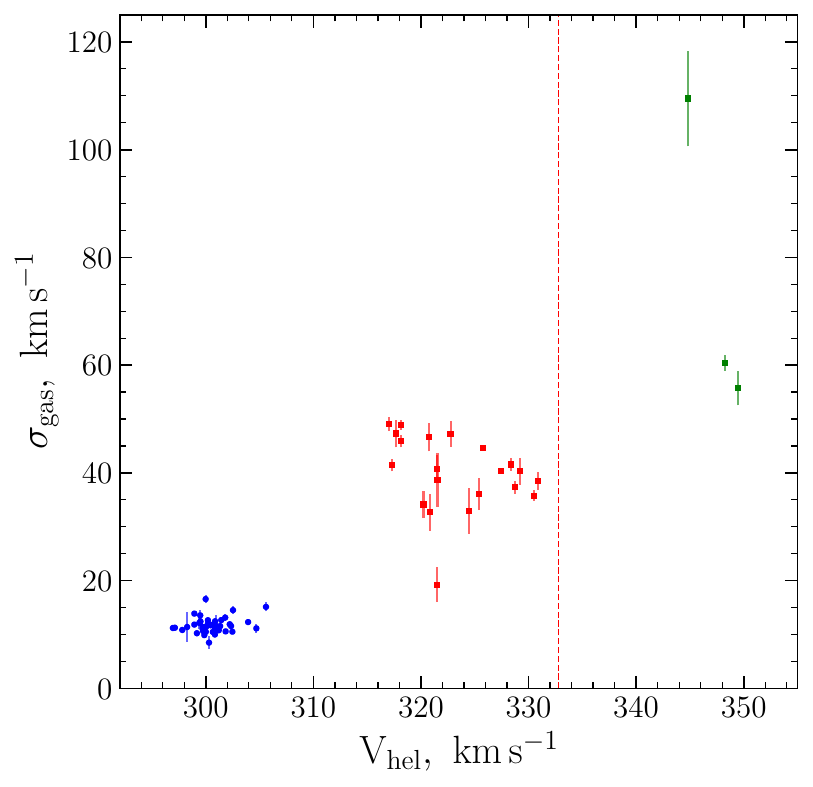}
    \caption{%
        The distribution of measured heliocentric velocities and
        gas dispersion for the narrow (blue circles),
        wide (red squares),
        and very wide components (green squares) of different emission lines
        measured during the \'echelle spectrum analysis.
        The very wide component of the [\ion{He}{ii}]~$\lambda$4686~\AA\ line, 
        has a gas dispersion, $\sigma_{gas}=393$~\kms,
        and therefore is not shown in the figure, but its velocity is indicated by the red dotted line.
        1$\sigma$ errors are shown with bars, but they are very small in the horizontal direction. 
        \label{fig:Vel_Vexp}}
\end{figure}

\begin{figure}
    \includegraphics[clip=,angle=0,width=0.48\textwidth]{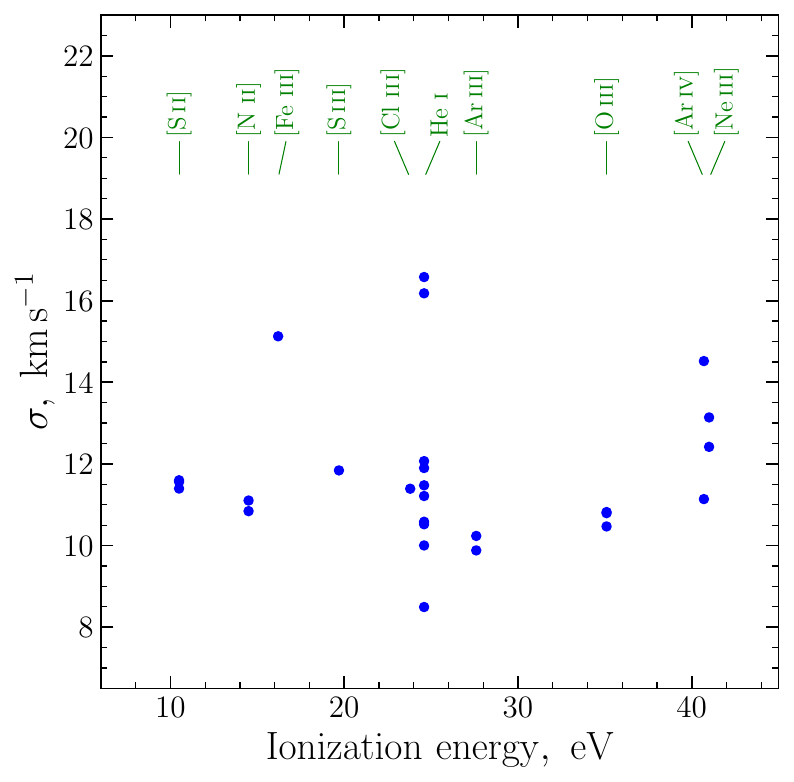}
    \includegraphics[clip=,angle=0,width=0.48\textwidth]{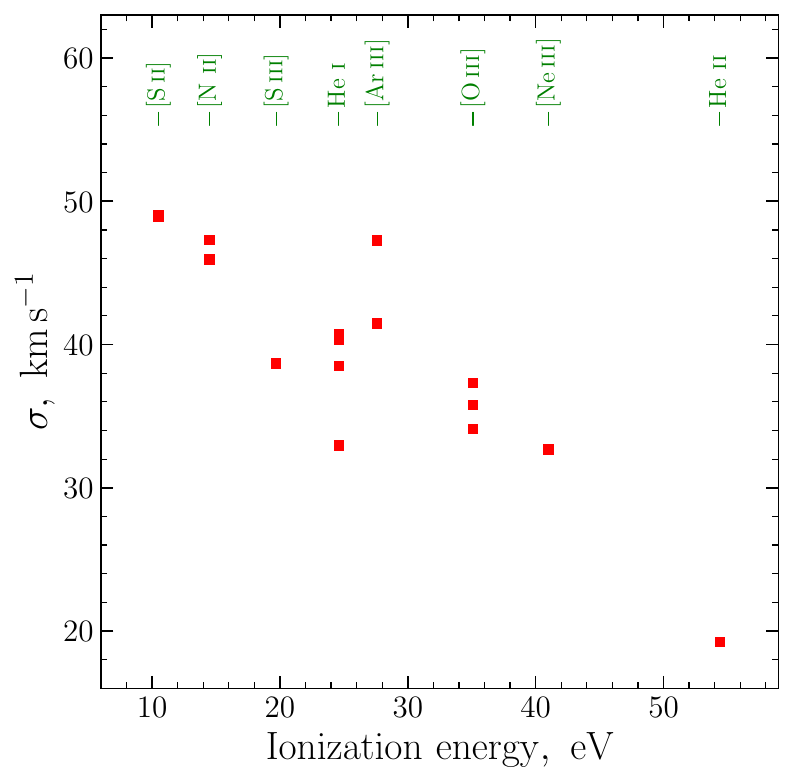}
    \caption{%
        The measured gas dispersion for different
        emission lines as a function of the consecutive ionization energy of
        atoms of different elements.
        The left panel shows this dependence for the NC,
        and the right panel for the BC.
        \label{fig:Ei_Vexp}}
\end{figure}

\subsection{Detected kinematic subsystems in the A2 region}
\label{txt:sub}

One of the potential features that the high spectral resolution of the \'echelle data provide
is an ability to find and study different kinematic subsystems
with different physical conditions (e.g., velocity dispersion of ionised gas and/or abundances of
chemical elements) that can be seen in the studied region.

In the case of the A2 region of the \IC\ galaxy, the \'echelle data immediately showed 
the presence of three kinematic subsystems
separated in velocity as well as by different velocity dispersions in the ionised gas
(hereafter in the text -- simply ``gas dispersion'').
All lines in the spectrum of region A2 with intensities greater  
than $\approx 3\%$ of I(H$\beta$)
allow us with certainty to detect the presence of two components, which in the following text
will be referred to as the ``narrow component'' (NC) and the ``broad component'' (BC).
The measured fluxes of the NC are about 5--7 times larger than those of the BC.
An example of such components is shown in Figure~\ref{fig:dual},
where the spectral region of the H$\alpha$ and nitrogen [\ion{N}{ii}]~$\lambda\lambda$6548,6584~\AA\ lines
is shown in the left plot 
and of the sulphur doublet [\ion{S}{ii}]~$\lambda\lambda$6716,6731~\AA\
in the right plot.

The brightest emission lines, H$\alpha$ and forbidden oxygen lines
[\ion{O}{iii}]~$\lambda$$\lambda$4959,5007~\AA, also show the presence of a third,
``very broad component'' (VBC hereafter; see left panel of Figure~\ref{fig:dual}),
which is about 5--7 times weaker than the BC.
Additionally, the VBC is found in the [\ion{He}{ii}]~$\lambda$4686~\AA\ line,
and this fact will be discussed in Section~\ref{txt:disc}.

In order to separate these components, 
each spectral line was fitted with two Gaussians (see Section~\ref{txt:analysis})
and no additional constraint on the decomposition was imposed.
The result of this decomposition is given in Tables~\ref{tab:Lines_kin} and~\ref{tab:Lines_kin1}.
The decomposition to components is presented only for those lines where the flux error
of the BC is less than the measured BC flux itself, otherwise only the NC was considered to be visible.
The H$\alpha$ and [\ion{O}{iii}]~$\lambda$$\lambda$4959,5007~\AA\ lines were fitted
with three components each.
The measured heliocentric velocities for all components of each line
and the gas dispersion ($\rm \sigma_{gas}$) for these components are shown 
in Tables~\ref{tab:Lines_kin} and~\ref{tab:Lines_kin1} as well.
The gas dispersion was recalculated from the measured FWHM of each component using the equation:
\begin{equation}
    \label{eq:exp}
    {\rm \sigma_{gas} = 0.43 \sqrt{V^2_{FWHM} - \Delta V^2_{therm} - \Delta V^2_{instr}}}
\end{equation}
where $\rm \Delta V_{instr}$ is the velocity correction recalculated from the instrumental FWHM correction
(equations~\ref{eq:blue} and \ref{eq:red} from this paper),
and $\rm \Delta V_{therm}$ is the correction for temperature broadening,
which was calculated as in \citet{1980afcp.book.....L}:
\begin{equation}
    \label{eq:exp1}
    {\rm \Delta V_{therm} = 21.4 \sqrt{\frac{T_e(ion)}{10^4 \cdot A(ion)}}}
\end{equation}
where $\rm T_{\rm e}(ion)$ is the electron temperature of the corresponding ion from
Table~\ref{t:Chem1}, and A(ion) is the atomic weight of that ion.
Since the measurements of the emission line centre have a very small 
formal error (only some meters per second),
they are not given in Table~\ref{tab:Lines_kin}, 
and the characteristic accuracy  of radial velocity determination
is taken to be the accuracy of the dispersion curve 
for LR \'echelle spectra, 300--400~\ms.

Figure~\ref{fig:Vel_Vexp} shows the distribution of heliocentric velocities versus gas dispersion
for the NC (blue circles), BC (red squares) and VBC (green squares) for different spectral lines.
We see that the distribution of heliocentric velocities and gas dispersions shows 
the presence of two, well separated, kinematic subsystems: the NC subsystem and the BC subsystem.
The mean heliocentric velocity of the NC subsystem is $\rm V_{hel} = 300.05\pm2.08$~\kms,
and for the BC subsystem is $\rm V_{hel} = 323.25\pm4.47$~\kms.
The mean gas dispersion of the NC subsystem is $\rm \sigma_{gas} = 11.90\pm2.44$~\kms,
and for the BC subsystem is $\rm \sigma_{gas} = 39.97\pm6.85$~\kms.
It can be assumed that the real accuracy of the mean heliocentric velocity
is better than the above, and an additional scatter is due to the fact that
lines of different elements have slightly different heliocentric velocities.
For example, the mean heliocentric velocity of the NC subsystem  measured 
with lines of the Balmer and Paschen series only is
$\rm V_{hel} = 300.38\pm0.80$~\kms\ with a mean gas dispersion $\rm \sigma_{gas}= 11.60\pm0.71$~\kms.
A similar picture can be seen for the VBC, where both gas dispersions and heliocentric velocities
are slightly different for hydrogen and oxygen but they are very close for both oxygen lines.

Figure~\ref{fig:Ei_Vexp} shows the dependence of the measured gas dispersion for different
emission lines as a function of the consecutive ionisation energy of atoms of these elements.
The left panel shows this dependence for the NC of emission lines,
and the right panel for the BC of emission lines.
The idea behind this picture is that in the case of a central ionising source that is
losing (has lost) some of its mass in the form of a stellar wind or of dropped shells,
and because of the stratification of the ionised gas region around this source
(as in the case of planetary nebulae),
the outer, colder regions of the nebula (\ion{H}{ii} region) expand
with higher velocities \citep[e.g.,][]{2006RMxAA..42...53M,2008ApJ...689..203R,2013A&A...558A..78J}.

The measured fluxes of the emission line components presented in Table~\ref{tab:Lines_kin},
were used to calculate abundances of chemical elements, which are shown
in columns (5) and (6) of Table~\ref{t:Chem1}.
For uniformity, the cold zone temperature was calculated
using the approximated temperature $T_{\rm e}$(OII) in both cases.
The comparison shows that the calculated abundances of oxygen and nitrogen for the BC  
are higher than the abundances calculated for the NC 
and this difference is about $3.3\sigma_{el}$ in oxygen and
$4.3\sigma_{el}$ in nitrogen.
For sulphur and argon the calculated abundances for the BC are higher than those for the NC
as well but these differences are less than $1\sigma_{el}$ due to
very large errors.


\section{Discussion}
\label{txt:disc}

An \'echelle spectrum of the \ion{H}{ii} region A2 in the irregular galaxy \IC\ 
shows the presence of three
spectral components (NC, BC and VBC), which differ in:
(1) the mean velocity dispersion,
(2) the average heliocentric velocity,
(3) the integral flux of components in each line,
(4) the abundance of chemical elements (for NC and BC only), and
(5) the behaviour of the measured gas dispersion for the different
emission lines as a function of the sequential ionization energy (for NC and BC only).

Additionally, (1) it can be argued that both the BC and the VBC of the
[\ion{He}{ii}]~$\lambda$4686~\AA\ line
in terms of velocities belong to the region of the BC, since the difference in velocity for the NC
[\ion{He}{ii}]~$\lambda$4686~\AA\ line (332.78~\kms)
and the average BC velocity (323.25~\kms) is only 2.1$\sigma_{el}$;
(2) the strong line ratios used in the BPT diagrams \citep{1981PASP...93....5B}
for both components (NC: log([\ion{O}{iii}]~$\lambda 5007/H\beta) = 0.72\pm0.01$,
log([\ion{N}{ii}]~$\lambda 6584/H\alpha) = -1.65\pm0.02$;
WC: log([\ion{O}{iii}]~$\lambda 5007/H\beta) = 0.78\pm0.03$,
log([\ion{N}{ii}]~$\lambda 6584/H\alpha) = -1.52\pm0.03$)
show that we are practically dealing with pure photoionization excitation.

Altogether, it can be assumed that the region generating the NC
in the \'echelle spectrum is the main body of the \ion{H}{ii} region A2, 
whose gas is well mixed,
and the recorded gas dispersion mainly characterizes the turbulent velocity of this gas
since its $\sigma_{gas}$ looks practically constant in the left panel of Figure~\ref{fig:Ei_Vexp}.    
A detailed study of \ion{H}{i} in \IC\ \citep{2010MNRAS.407..113V}
found the systemic velocity of the galaxy to be $V_{sys}$(\ion{H}{i}) = 302~\kms,
which agrees very well with the mean velocity of the NC region found from our data.
The oxygen abundance we calculated for the NC (see Table~\ref{t:Chem1}) also
agrees very well with the oxygen abundance for the same region
from \citet{2009A&A...499..455C} (region \#3 with O/H=8.06$\pm$0.03~dex).

The BC is apparently formed in a separate region around young, hot stars (or just a single hot massive star),
the wind from which ``injects'' mechanical energy into the surrounding space
and encourages the expansion of the jettisoned matter enriched with chemical elements.
This hot massive star (stars) is a central ionising source for this region as well. 
For the BC region, the contribution of turbulent gas velocity is probably smaller than that of  
expansion and the expansion velocity is close to the $\sigma_{gas}$ 
of the outermost region, $\sim 50$~\kms\ (bottom panel of Figure~\ref{fig:Ei_Vexp}).

The presence of the VBC of the \ion{He}{ii}~$\lambda$4686~\AA\ line indicates
that this region is most likely associated with the presence of a Wolf-Rayet star (WR hereafter).
\citet{2009A&A...499..455C} studied the \ion{H}{ii} regions A1 and A2 of the \IC\ galaxy
using images from different filters on the Hubble Space Telescope and
the VLT telescope. In their study, they found six WR stars
in these two \ion{H}{ii} regions and showed their positions.
Comparison of the position of the \'echelle object fibre during the SALT observations
(see bottom panel of Figure~\ref{fig:IC4662_Ha} of this paper) 
and the positions of WR stars detected by \citet{2009A&A...499..455C}
(see Figure~2 of that work) shows that the HRS fibre
was placed on that part of the A2 region which contains WR star A2--WR2.
The spectral aperture \#3 of \citet{2009A&A...499..455C} included two WR stars,
A2--WR2 and A2--WR3, and these authors suggested that one of those stars could be a WC star
and the other a WN5--6 star.
Our spectral \'echelle data do not show any additional lines of the WR star
other than the \ion{He}{ii}~$\lambda$4686~\AA\ line.
However, it should be noted here that the level of accumulated signal in the spectral region
$\sim 5800$~\AA\ is much higher than in the blue 4600--4700~\AA\ region
because of the overall sensitivity of the spectrograph.
{ Since there is no indication of the presence of a broad \ion{C}{iv}$\lambda5801-12$ line in the \'echelle spectrum, the absence of these spectral lines unequivocally confirms that the star A2–WR2 is of the WN type rather than the WC type. The measured FWHM = 14.31\AA\ for the emission line allows refinement of its ionization subclass to WN7–8 \citep{1996MNRAS.281..163S}. The expected accuracy of such classification is approximately $\pm1$ subclass with a probability of $\sim80\%$ \citep{2006A&A...457.1015H}. Additional confirmation of its classification as a late WN (WNL) star comes from the enhanced nitrogen content in the emission spectrum, which is characteristic of regions enriched by the ejecta of WNL stars.}

Most of the WR stars found in \citet{2009A&A...499..455C} are located on the edges of the \ion{H}{ii} regions, A1 and A2. It is known that most of the studied WR stars in our Galaxy are also located outside known stellar clusters. And since most massive stars are born in clusters \citep[e.g.][]{2007ARA&A..45..481Z} it is assumed that most of the WR stars studied have been ejected from parent clusters. Such stars could be ejected either by dynamic processes in the cluster core \citep{1967BOTT....4...86P,1986ApJS...61..419G} or as a result of a supernova explosion in a close system of two massive stars \citep{1961BAN....15..265B,1991AJ....102..333S}.
In both scenarios, the ejected WR star must be located close to the parent cluster, since the precursor WR star is a very massive star with a mass of at least 25~M$_\odot$, having a lifetime of at most 5~Myr. Detailed studies of individual WR stars in our Galaxy find their inferred mother clusters based on the observed radial velocities of these stars and their proper motions \citep[e.g.][]{2009MNRAS.400..524G,2010MNRAS.403..760G,2013MNRAS.429.3305B}. { Therefore, the velocity of the NC from our study can naturally be interpreted as the velocity of the parent \ion{H}{ii} region A2, which closely matches the systemic velocity of the galaxy. In this framework, the velocity of the BC is interpreted as the velocity of the circumstellar envelope surrounding the WN star (as well as the velocity of star itself). The velocity difference, $dV = (V_{hel}^{BC} - V_{hel}^{NC}) = 23$~km/s, represents the radial component of the ejection velocity from the parent cluster located within \ion{H}{ii} region A2. This interpretation is further supported by the location of the WN star at the edge of \ion{H}{ii} region A2 (see the right panel of Figure\ref{fig:IC4662_Ha}).}
    
Modern studies of the evolution of WR stars and their circumstellar envelopes distinguish two types of envelope. In the first case, these are compact envelopes of very young WR stars of WNL type, where the stellar wind is still confined to the region occupied by circumstellar matter dumped at a previous stage of red supergiant (RSG) or LBV evolution and therefore show the signatures of CNO-processing \citep[e.g,][]{1981ApJ...249..195C,1981NInfo..49...21L,2000AJ....120.2670G,2010MNRAS.405..520G}. On the other hand, winds from more evolved WR stars of WNE type interact directly with the local interstellar medium and create more extended envelopes, whose abundances of chemical elements are equal to the abundances of the surrounding \ion{H}{ii} regions \citep[e.g.,][]{1992A&A...259..629E,2011MNRAS.418.2532S}. Our analysis shows that the abundance of the chemical elements, oxygen and nitrogen, in the BC region is systematically higher than the abundance of the same elements in the NC region, which is the main body of the A2 region. This result also fits logically into the picture of ejected stellar material around a very young WN star where a fast wind has caught up with material enriched with fusion products previously shed by the RSG, and contributes to its expansion. Since its ionisation subclass is WN7--8 it is quite possible that the fast WR wind is still just passing through the zone created by the slow RSG wind \citep{2009MNRAS.400..524G}.

{
Let us now support the proposed ideas with various estimates based on existing models:\\
(1) According to the models of \citet{1977ApJ...218..377W} and \citet{1985Natur.317...44C}, the expansion velocity of a shell created by the wind of a massive star can be estimated as $v_{exp} \propto (L_w/n)^{(1/5)}$, where $L_w$ is the mechanical luminosity of the wind, and $n$ is the density of the surrounding medium. The mechanical luminosity of the wind can be calculated using the formula $L_w = (1/2) \dot{M} v_{\infty}^2$, where $\dot{M}$ is the mass-loss rate, and $v_\infty$ is the terminal wind velocity. For WR stars of type WN7-8, these parameters can be taken from Table~1 of \citet{2007ARA&A..45..177C} as $\dot{M} = 5\times10^{-5} M_\odot/year$, $v_\infty = 1000$~\kms, leading to an estimated mechanical wind luminosity of $L_w  \approx 10^{36}-10^{37}$~erg/s. Using the measured electron density $n_e \approx 100$~cm\(^{-3}\), the theoretical expansion velocity is expected to be $\sim30-60$~\kms, which is in good agreement with the measured velocity dispersion of the SC $\approx 40$~\kms.\\
(2) Since the size of the studied giant \ion{H}{ii} region is approximately 25~pc, using the average expansion velocity, the age of the shell can be estimated as $ t_{age} \approx R_{shell} / v_{exp} \approx 12.5/40 \approx 3\times10^5$~years, which is consistent with the short evolutionary phase of WR stars following the red supergiant (RSG) stage.\\
(3) For a WN7-8 star, the typical terminal wind velocity is $v_\infty \approx 1000-1200$~\kms\ \citep{2006A&A...457.1015H,2007ARA&A..45..177C}. Then, the ratio $\sigma(HeII\,\lambda4686)/v_\infty \approx 0.3$, which is consistent with theoretical models of line formation in WR star winds \citep{1991A&A...247..455H}.\\ 
(4) Nitrogen enrichment (a difference with a significance of $4.4\sigma$ between NC and BC) quantitatively supports the assumption that the material was ejected during the previous RSG/LBV stage. The typical increase in nitrogen content in shells following the RSG phase is 0.2--0.4~dex \citep{2001ApJ...551..764L}, which corresponds to the observed difference of 0.22~dex between NC and BC.

\subsection{Very Broad Component}
\label{txt:VBC}

The VBC is observed in many irregular galaxies \citep[for example,][]{2020AstBu..75..361O,2020MNRAS.495.4347B,2021MNRAS.508.2650E} and is associated with the action of winds from massive stars. This component represents a unique "window" into the processes occurring in the immediate vicinity of the WR star. The extreme velocity dispersion of the VBC in the \ion{He}{ii} $\lambda$4686~\AA\ line is directly linked to the fast wind of the WR star and corresponds to approximately 1/3 of the expected terminal wind velocity for a WN7-8 type star (see above). Significantly lower, but still high dispersions of the VBC in the [\ion{O}{iii}]~$\lambda$5007~\AA\ and H$\alpha$ lines (see Table~\ref{tab:Lines_kin1}) indicate the presence of stratification in the wind acceleration region.

The three-component structure (VBC, BC, NC) is, in essence, a direct observation of different zones in the classical wind bubble model \citep{1977ApJ...218..377W}: the VBC in the \ion{He}{ii} line corresponds to the freely flowing wind zone, the VBC in the [\ion{O}{iii}]~$\lambda$5007~\AA\ and H$\alpha$ lines corresponds to the hot gas behind the shock front, whilst the BC corresponds to the compressed shell of cooled gas at the boundary with the surrounding interstellar medium. The relatively low intensity of the VBC (approximately 5--7 times weaker than the BC) is consistent with the small filling factor of hot gas in such a structure. The difference in radial velocities between the VBC components (Table~\ref{tab:Lines_kin1}) may indicate non-spherical expansion or the presence of asymmetry in the density distribution of the surrounding medium, which is consistent with modern 3D models of the interaction of WR star winds with a heterogeneous medium \citep{2011ApJ...734L..26V}.

It should be noted that the observation of the VBC with such characteristics in a low-metallicity environment provides valuable information about the influence of heavy element content on stellar wind properties and their interaction with the surrounding medium. In systems with low metallicity, a decrease in both the mass loss rate and terminal wind velocity of WR stars is expected \citep{2005A&A...442..587V}. The presented observations show that even under such conditions, stellar wind is capable of creating a complex dynamic structure in the surrounding environment, which has important implications for understanding the feedback of massive stars in the early Universe.

\subsection{Implications for galaxy evolution}
\label{txt:implication}

In general, observations of the \ion{H}{ii} region A2 in the \IC\ galaxy provide important information about star formation processes and feedback in low-metallicity environments:

(1) The detection of significant differences in chemical composition (0.15-0.20~dex) between kinematic subsystems on a scale of only 25~pc confirms that local enrichment plays a substantial role in the chemical evolution of dwarf galaxies, contrary to the widespread belief about rapid and complete mixing of the interstellar medium in these systems. In \citet{2005AJ....130.1558K}, the authors indicated that despite the established conviction that the interstellar medium in irregular galaxies is well mixed, very often information about the average metallicity of such galaxies is known with an accuracy of up to 0.3--0.4~dex, which is related both to possible processes of local enrichment and an insufficient number of quality observations.

(2) The identification of an expanding shell around the WR star with increased nitrogen content demonstrates ``in action'' the mechanism of interstellar medium enrichment with products of massive star evolution. This is especially important for understanding the early stages of the chemical evolution of the Universe, when similar processes could dominate in primordial galaxies.

(3) The effective mechanical impact of the WR star wind on the surrounding environment even in low-metallicity conditions indicates that stellar feedback may be a key factor in regulating star formation in metal-poor systems. The expanding shell not only distributes enriched material, but also may create conditions for the triggered formation of the next generation of stars.

(4) The discovery of the complex kinematic structure of the \ion{H}{ii} region A2 with three different components emphasises the need to develop multiphase models of the interstellar medium in low-metallicity galaxies for adequate interpretation of observations and understanding of feedback cycles between star formation and interstellar medium properties.
}

\section{Conclusions}
\label{txt:summ}

The bright \ion{H}{ii} region A2 of the irregular  galaxy \IC\
was studied using long-slit and \'echelle spectroscopy
at the SALT telescope. This work has shown that:

1. The calculated abundances of chemical elements obtained from different spectral data
are consistent with each other within the available errors,
and hence the spectral data obtained from the HRS \'echelle spectrograph,
corrected for the spectral sensitivity curve,
can be used to determine the abundances of chemical elements of emission nebulae.

2. The analysis of \'echelle data revealed three spectral components,
apparently belonging to different subsystems of the investigated \ion{H}{ii} region.
Their physical and kinematic characteristics were compared, as were the abundances 
of the chemical elements, oxygen, nitrogen, sulphur and argon,
which were calculated using the T$_{\rm e}$ method.

3. The analysis of \'echelle data as well as a comparison with previous studies of the region A2 in \IC\ allowed us to suggest that one of the detected subsystems belongs to the region around a hot young WR star of WNL type, which has been ejected from the parent cluster located in the main body of region A2. The wind from this WR star ``injects'' mechanical energy into the surrounding area and contributes to the expansion of the envelope shed in a previous stage of evolution (red supergiant or LBV) and enriched with chemical elements. This assumption is very well supported by comparisons with theoretical models.

\begin{acknowledgements}
This work is based on observations obtained with the Southern African Large Telescope (SALT),
program 2017-1-MLT-001 (PI: Kniazev).
A.Y. acknowledges support from the National Research Foundation (NRF) of South Africa.
{ The author expresses gratitude to the anonymous reviewer for comments and suggestions, which helped to improve the description of the obtained results and their interpretation.}
\end{acknowledgements}

\bibliographystyle{raa}
\bibliography{IC4662}

\label{lastpage}
\end{document}